\documentclass[12pt]{article}
\def\theequation{\arabic{section}.\arabic{equation}}
\usepackage{amssymb}

\newcounter{rown}

\begin{document}
\renewcommand{\thefootnote}{\fnsymbol{footnote}}
\renewcommand{\theequation}{\thesection.\arabic{equation}}
\title{Classification of irreps and invariants of the $N$-extended Supersymmetric Quantum Mechanics.}
\author{Z. Kuznetsova${}^{a}$\thanks{{\em e-mail: zhanna@cbpf.br}}~, M. Rojas${}^{b}$\thanks{{\em e-mail: 
mrojas@if.ufrj.br}} 
and F. Toppan${}^{c}$\thanks{{\em e-mail: toppan@cbpf.br}}
\\ \\
${}^a${\it Dep. de F\'{\i}sica, Universidade Estadual de Londrina,}\\{\it Caixa Postal 6001, Londrina (PR), Brazil}\\ 
${}^b${\it Inst. de F\'{\i}sica, Universidade Federal do Rio de Janeiro,}\\
{\it Caixa Postal 68528, cep 21945-910, Rio de Janeiro (RJ), Brazil}\\ 
${}^c${\it CBPF, Rua Dr.}
{\it Xavier Sigaud 150,}
 \\ {\it cep 22290-180, Rio de Janeiro (RJ), Brazil}}
\maketitle
\begin{abstract}
We present an algorithmic classification of the irreps of the $N$-extended
one-dimensional supersymmetry algebra linearly realized on a finite number of fields. Our work is based on the $1$-to-$1$ \cite{pt} correspondence between Weyl-type Clifford algebras (whose irreps are fully classified) and classes of irreps of the $N$-extended $1D$ supersymmetry.
The complete classification of irreps is presented up to $N\leq 10$. The fields of an irrep are accommodated
in $l$ different spin states. $N=10$ is the minimal value admitting length $l>4$ irreps. The classification
of length-$4$ irreps of the $N=12$ and {\em real} $N=11$ extended supersymmetries is also explicitly
presented.\par  
Tensoring irreps allows us to systematically construct manifestly ($N$-extended) supersymmetric multi-linear invariants
{\em without} introducing a superspace formalism. Multi-linear invariants can be constructed both for {\em unconstrained} 
and {\em multi-linearly constrained} fields. A whole class of off-shell invariant actions are produced
in association with each irreducible representation. The explicit example of the $N=8$ off-shell action of the
$(1,8,7)$ multiplet is presented.\par
Tensoring zero-energy irreps leads us to the notion of the {\em fusion algebra} of the $1D$ $N$-extended supersymmetric vacua.

\end{abstract}
\vfill 
\rightline{CBPF-NF-010/05}

\newpage
\section{Introduction}

Supersymmetric Quantum Mechanics is a twenty-five years old topic \cite{wit} which is still under intensive development
and even received in the last few years a considerable renewed attention. Part of the reason is due to the wide range
of applicability of one-dimensional supersymmetric theories and especially superconformal quantum mechanics \cite{sca}
for extremal black holes \cite{bh}, in the AdS-CFT correspondence \cite{ads} (when setting $AdS_2$), for investigating partial breaking of extended supersymmetries \cite{{ikp},{spbr}} and so on. Another very important motivation
is due to the fact that considerable advances in understanding the structure of large-$N$ extended supersymmetry itself
have been made in the recent years.  It is well known that large $N$ (up to $N=32$, starting from the maximal, eleven-dimensional supergravity) one-dimensional supersymmetric quantum mechanical models are automatically derived \cite{dire} from
the dimensional reduction of higher-dimensional supersymmetric field theories. Large $N$ one-dimensional supersymmetry
on the other hand (possibly in the $N\rightarrow \infty$ limit) even emerges in condensed matter phenomena, as described by the BCS model, see e.g. \cite{int}. \par
Controlling one-dimensional $N$-extended supersymmetry for arbitrary values of $N$ (that is, the nature of its representation theory, how to construct manifestly supersymmetric invariants, etc.)
is a technical, but challenging program with important consequences in many areas of physics, see e.g. the discussion in \cite{glpr} concerning the nature of {\em on-shell versus off-shell} representations,
for its implications in the context of the supersymmetric unification of interactions. \par
Along the years, progresses came from two lines of attack. In the pivotal work of \cite{dr} irreducible representations 
were investigated to analyze supersymmetric quantum mechanics. A special role played by Clifford algebra was pointed
out \cite{brr}. Clifford algebras were also used in \cite{gr} to construct representations of the extended one-dimensional
supersymmetry algebra for arbitrarily large values of $N$. Another line of attack, consisted in using the superspace,
so that manifest invariants could be
constructed through superfields. For low values of $N$ this is indeed the most convenient approach. However,
with increasing $N$, the associated superfields are getting highly reducible and require the introduction of constraints
to extract the irreducible representations. This approach gets soon unpractical for large $N$. Indeed, only very recently
a manifestly $N=8$ superfield formalism for one-dimensional theory has been introduced, see \cite{abc} and references therein. A manifest superfield formalism is however still not available for larger values of $N$.\par
In \cite{pt} a contribution in understanding the nature of the linearly realized irreducible representations of arbitrary $N$-extended supersymmetries was made,
proving that {\em any} irrep can be classified and recovered from an associated Clifford algebra. This is the starting point 
for the present work.  \par
In this paper we furnish a systematic classification of the irreps of the $N$-extended one-dimensional 
supersymmetry algebra. In \cite{pt} it was shown that
all such irreps fall into classes of equivalence in $1$-to-$1$ correspondence with a certain subclass
(the Weyl subclass) of Clifford algebras, the dimensionality of the Clifford algebra being linked to the
integer $N$ labeling the extension of the supersymmetry. It was further proven
that any given irrep can be constructed by applying a so-called {\em dressing transformation}
to the length-$2$ irrep (the length of an irrep expresses the number of its different spin states)
belonging to its equivalence class.
The classification of length-$2$ irreps is immediately available, being borrowed from the known \cite{{abs},{por},{oku}} classification 
of Clifford algebras. On the other hand, the classification of irreps of general length $l>2$ requires the investigation of the
properties (mostly the {\em locality} property, discussed in Section {\bf 2}) of the dressing transformations. 
In \cite{pt} length-$3$ irreps were easily classified, but no general attempt was made to classify irreps
with length $l\geq 4$ (only one specific example, the unique length-$4$ irrep of the $N=3$ supersymmetry which,
btw, coincides with the $N=3$ ``enveloping" representation, as discussed in Section {\bf 5}, was explicitly constructed). 
The full classification of general length irreps can be achieved and systematically organized by using specific properties
of the Clifford irreps (encoded, e.g., in the so-called ``block-symbols" of Section {\bf 4}) which can be algorithmically
computed according to the method presented in \cite{crt2}. In our approach an important role is
also played by the notion of {\em oxidized} Clifford algebras
(corresponding to, essentially, the maximal number of Clifford generators which can be accommodated into irreducible representations of a given
matrix size, see \cite{kt}), together with their associated {\em oxidized} extended supersymmetries discussed in Section {\bf 3}. The fact that oxidized Clifford algebras are either real or quaternionic \cite{kt} can be used, e.g., to construct,
for $N=3,5 \quad mod \quad 8$, two separated classes of irreps, real, denoted with ``$^{(\ast)}$", and quaternionic,
denoted with ``$^{(\ast\ast)}$" (the $mod\quad 8$
property is in consequence of the Bott's periodicity of Clifford Gamma matrices; the remaining values of $N$ admit a unique
type of irreps).\par
The algorithmic \cite{crt2} presentation of Clifford irreps allows us to classify, for any given $N$, the irreducible representations of the one-dimensional supersymmetry algebra (\ref{susyalg}) realized on a {\em finite} number of fields
and to explicitly construct a representative in each irreducible class. In this paper we limit ourselves to
explicitly present the complete
classification of irreps for $N\leq 10$ (and furnish the classification of length-$4$ irreps for the oxidized
$N=11^{(\ast)},12$ supersymmetries).\footnote{It is worth pointing out that our method can be applied to arbitrarily large values of $N$. It should be taken into account however that, while some properties of the irreps, like their dimensionality (\ref{irrepdim}) or the fact that the class of irreps is closed
under the {\em high $\Leftrightarrow$ low} spin duality (\ref{hiloduality}), can be easily stated, at increasing  
$N$ not only the actual computations, but also the presentations of the complete lists of results are getting more and more cumbersome.}  \par
It deserves to be mentioned at this point that the inequivalent irreducible representations have been used in the literature
to produce super-particle models moving in one and higher dimensional target manifolds. The three length-$3$ irreps of the
$N=4$ supersymmetry (namely $(1,4,3)$, $(2,4,2)$ and $(3,4,1)$, see Appendix {\bf A}) were respectively used, e.g., to
construct one-dimensional \cite{{ikp},{one},{onebis}}, two-dimensional \cite{two} and three-dimensional \cite{three} Supersymmetric Quantum Mechanics. An updated list of references concerning the Supersymmetric Quantum Mechanics constructed via the length-$3$ $N=8$ irreps can be found in \cite{abc}.
\par
Besides classifying irreps, in this paper we also point out that tensoring irreps allows us to systematically construct
manifestly (multi)-linear invariants of the $N$-extended supersymmetry algebra (\ref{susyalg}). The reason lies in the fact
that the component fields of highest spin in the tensored multiplets transform, under supersymmetry, as time-derivatives.
They can therefore be used as lagrangian terms entering a manifest invariant action. It is worth mentioning that in this framework these invariants are constructed {\em without} introducing the notion of superspace and of their associated superfields. As already recalled, for large values of $N$, superfields are of limited use.  \par
In our framework two big classes of invariants, {\em constrained} and {\em unconstrained}, can be constructed. Indeed, the tensor product of irreps produces, in general,
reducible representations. A basic illustrative example can be considered the $N=4$ self-tensoring of the $(1,4,3)$ multiplet,
see Section {\bf 6}, producing at the leading order the $N=4$ enveloping representation (the enveloping
representations are reducible for $N\geq 4$). Therefore, either we extract the invariants in terms of unconstrained fields from
the highest spin component(s) of the tensored reducible representations or we implement bilinear (in general, multilinear
for multiple tensorings) constraints to extract an irreducible representation realized with bilinear (multilinear) combinations
of the original fields. The highest spin components of the bilinearly (multilinearly) realized irrep generate invariants, once
the bilinear (multinear) constraints are taking into account.\par
We further discuss the possibility to accommodate, within our framework, manifest $N$-extended invariants for $\sigma$-model types of terms \cite{sigma}.\par
A whole class of off-shell invariants for arbitrarily large values of $N$ is introduced and discussed in Section {\bf 7}. 
For any given irrep, it is obtained by ``embedding" the most general function of the fields entering the irrep into the
``enveloping representation" (discussed in Section {\bf 5}) of the $N$-extended supersymmetry. Let $f$ denote such a general function, an 
invariant of the $N$-extended supersymmetry is given by the integral $\int dt \left( Q_1\ldots Q_N  f\right)$.
For $N=4$ the off-shell invariant actions have the ``right" dimension of the ordinary kinetic term. Within
our construction we are able to produce the most
general off-shell invariant action of ``right dimension" for the $(1,8,7)$ multiplet of the $N=8$ supersymmetry, by ``covariantizing" the
$N=4$ ($1,4,3$) case w.r.t. the octonionic structure constants entering the Clifford algebra $Cl(0,7)$.
\par
Another concept introduced in this work, is that of the {\em fusion algebra} of the zero-energy irreps (i.e. of the supersymmetric vacua) of the $N$-extended one-dimensional supersymmetry. 
This concept is mimicked after the corresponding notion for RCFT's, see \cite{gab}. It allows us to encode, in integer-valued
fusion matrices, the decomposition into irreps of the tensor products of the supersymmetry irreps at zero energy.  
\par
The plan of the paper is the following. In Section {\bf 2} the \cite{pt} $1$-to-$1$ connection between
Weyl-type Clifford algebras and classes of irreps of the $N$-extended supersymmetry algebra is reviewed.
In Section {\bf 3} the classification of Clifford algebras is recalled. The notion of {\em oxidized} Clifford
algebras, leading to {\em oxidized} and {\em reduced} $N$-extended supersymmetries, is introduced.
In Section {\bf 4} we use the knowledge of Clifford algebras (encoded in the so-called ``block-symbols") to
classify (and explicitly construct representatives for each irreducible multiplet) arbitrary length-$l$ irreps
of the $N$-extended supersymmetry. The complete classification of irreps is explicitly reported up to $N\leq 10$. In Section {\bf 5} 
some comments are made on the nature of the enveloping representations. In Section {\bf 6}, the tensorings of irreps
and their decomposition into irreps is used to construct multi-linear invariants of the $N$-extended supersymmetries.
These invariants can be used as potential, constant kinetic, $\sigma$-model type, etc., terms entering a manifest
$N$-extended invariant action, {\em without} introducing a superspace and the associated
superfield formalism. The decomposition
of tensored-products into irreps leads us to introduce the two big classes of the {\em unconstrained} invariants and the
bilinearly (in general, multilinearly) {\em constrained} invariants, see the $(1,4,3)\otimes (1,4,3)$ $N=4$
example. 
In Section {\bf 7} the class of ``enveloping" off-shell invariants produced by any given irrep of the $N$-extended supersymmetry
are discussed.
In Section {\bf 8} the notion of the fusion algebra of the supersymmetric vacua is introduced.
Non-negative integral-valued fusion matrices encode the tensoring of the zero-energy supersymmetry irreps.
The associativity of the tensoring implies the commutativity of the fusion matrices. 
In the Conclusions we present a more detailed analysis of the results here achieved and discuss future perspectives.
The work is further integrated with four appendices. Appendix {\bf A} is devoted to explicitly present 
representatives of each supersymmetry irrep for all $N$-extended supersymmetries with $N\leq 8$. In Appendix
{\bf B} we furnish the complete classification of the irreps for the $N=9,10$ extended supersymmetries and
explicitly present the length-$4$ classification of irreps for the oxidized $N=11^{(\ast)},12$ extended
supersymmetries. In Appendix {\bf C} the tensoring of the $N=2$ irreps and of some selected examples of the $N=4$
irreps, is explicitly presented. In Appendix {\bf D} we produce the fusion matrices (for both cases, 
either disregarding or taking into account the statistics of the multiplets) of the $N=2$ supersymmetric vacua.

\setcounter{equation}{0}
\section{Irreps of the $N$-extended $d=1$ supersymmetry and Clifford algebras: the connection revisited}

In this section we review the main results of ref. \cite{pt} concerning the classification of irreps of the
$N$-extended one-dimensional supersymmetry algebra. \par
The $N$ extended $D=1$ supersymmetry algebra is given by
\begin{eqnarray}\label{susyalg}
\{ Q_i,Q_j\}&=& \eta_{ij} H
\end{eqnarray}
where the $Q_i$'s are the supersymmetry generators (for $i,j=1,\ldots , N$) and $H\equiv -i\frac{\partial}{\partial t}$ is a
hamiltonian operator ($t$ is the time coordinate). If the diagonal matrix $\eta_{ij}$ is pseudo-Euclidean (with signature
$(p,q)$, $N=p+q$) we can speak of generalized supersymmetries. The analysis of \cite{pt} was done for this general case.
For convenience in the present paper (despite the fact that our results can be straightforwardly generalized to
pseudo-Euclidean supersymmetries, having applicability, e.g., to supersymmetric spinning particles moving in pseudo-Euclidean
manifolds) we work exclusively with ordinary $N$-extended supersymmetries. Therefore for our purposes here $\eta_{ij}\equiv \delta_{ij}$.\par 
The ($D$-modules)
representations of the (\ref{susyalg}) supersymmetry algebra realized in terms of {\em linear}
transformations acting on {\em finite} multiplets of fields satisfy the following properties. The total number of bosonic fields equal the total number of
fermionic fields. For irreps of the $N$-extended supersymmetry the number of bosonic (fermionic) fields is given by
$d$, with $N$ and $d$ linked through
\begin{eqnarray}\label{irrepdim}
N&=& 8l+n,\nonumber\\
d&=& 2^{4l}G(n),
\end{eqnarray}
where $l=0,1,2,\ldots$ and $n=1,2,3,4,5,6,7,8$.
$G(n)$ appearing in (\ref{irrepdim}) is the Radon-Hurwitz function \cite{pt}
{{
{ {{\begin{eqnarray}&\label{radonhur}
\begin{tabular}{|c|cccccccc|}\hline
$n  $&$1$&$2$&$3$& $4$&$5$&$6$&$7$&$8$\\ \hline
$G(n)  $&$1$&$2$&$4$& $4$&$8$&$8$&$8$&$8$\\ \hline
\end{tabular}&\nonumber\\
&&\end{eqnarray}}} }  
}}
The modulo $8$ property of the irreps of the $N$-extended supersymmetry is in consequence of the famous modulo
$8$ property of Clifford algebras. The connection between supersymmetry irreps and Clifford algebras is specified
later.\par
Due to the fact that the $D=1$ dimensional reduction of the maximal $N=8$ supergravity
produces a supersymmetric quantum mechanical system with $N=32$ extended number of supersymmetries,
it is convenient to explicitly report the number of bosonic/fermionic component fields in any given irrep of
(\ref{susyalg}) for any $N$ up to $N=32$. We get the table
{{
{ {{\begin{eqnarray}&\label{N32}
\begin{tabular}{|ll|ll|ll|ll|}\hline
$N=1  $&$1$&$N=9$&$16$& $N=17$&$256$&$N=25$&$4096$\\ \hline
$N=2  $&$2$&$N=10$&$32$&$N=18$&$512$&$N=26$&$8192$\\ \hline
$N=3  $&$4$&$N=11$&$64$&$N=19$&$1024$&$N=27$&$16384$\\ 
$N=4  $&$4$&$N=12$&$64$&$N=20$&$1024$&$N=28$&$16384$\\ \hline
$N=5  $&$8$&$N=13$&$128$&$N=21$&$2048$&$N=29$&$32768$\\ 
$N=6  $&$8$&$N=14$&$128$&$N=22$&$2048$&$N=30$&$32768$\\ 
$N=7  $&$8$&$N=15$&$128$&$N=23$&$2048$&$N=31$&$32768$\\ 
$N=8  $&$8$&$N=16$&$128$&$N=24$&$2048$&$N=32$&$32768$\\ \hline
\end{tabular}&\nonumber\\
&&\end{eqnarray}}} }  
}}
The bosonic (fermionic) fields entering an irreducible multiplet can be grouped together according
to their dimensionality. Throughout this paper we use, interchangeably, the words ``dimension" or 
``spin" to refer to the dimensionality of the component fields. It is in fact useful, especially when
discussing the $D=1$ dimensional reduction of higher-dimensional supersymmetric theories, to refer at the
dimensionality of the $D=1$ fields as their ``spin". 
The number (equal to $l$) of different dimensions (i.e. the number of different spin states) of a given irrep, will be 
referred to as the {\em length} $l$ of the irrep. Since there are at least two different spin states 
(one for bosons, the other for fermions), obtained when all bosons (fermions) are grouped together within the same spin, 
the minimal length of an irrep
is $l=2$. \par
A general property of (linear) supersymmetry in any dimension is the fact that the states of highest spin in a given multiplet are auxiliary fields, whose supersymmetry transformations are given by total derivatives. Just for $D=1$ total derivatives coincide with the (unique) time derivative. Using this specific property of the one-dimensional supersymmetry
it was proven in \cite{pt} that all finite linear irreps of the (\ref{susyalg}) supersymmetry algebra fall into 
classes of equivalence, each class of equivalence being singled out by an associated minimal length ($l=2$) irreducible multiplet. It was further proven that the minimal length irreducible multiplets are in $1$-to-$1$ correspondence with a subclass of Clifford algebras (the ones which satisfy a Weyl property). The connection goes as follows. The supersymmetry generators
acting on a length-$2$ irreducible multiplet can be expressed as
\begin{eqnarray}\label{length2irrep}
Q_i&=& \frac{1}{\sqrt 2}\left( \begin{tabular}{cc} $0$& $\sigma_i$\\
${\widetilde \sigma}_i\cdot H$& $0$
\end{tabular}
\right) 
\end{eqnarray}
where the $\sigma_i$ and ${\widetilde\sigma}_i$ are matrices entering a Weyl type (i.e. block antidiagonal) 
irreducible representation of the Clifford algebra 
relation
\begin{eqnarray}\label{weylclifford}
\Gamma_i =\left( \begin{tabular}{cc} $0$& $\sigma_i$\\
${\widetilde \sigma}_i$& $0$
\end{tabular}
\right)\quad &,&\quad\{ \Gamma_i,\Gamma_j\}= 2\eta_{ij} 
\end{eqnarray}
The $Q_i$'s in (\ref{length2irrep}) are supermatrices with vanishing bosonic 
and non-vanishing fermionic blocks, acting on an irreducible multiplet $m$ 
(thought as a column vector) which can be either bosonic or fermionic\footnote{We conventionally 
consider
a length-$2$ irreducible multiplet as bosonic if its upper half part of component fields is bosonic 
and its lower half is
fermionic. It is fermionic in the converse case.}. 
The connection between Clifford algebra irreps of Weyl type and minimal length irreps of the $N$-extended 
one-dimensional supersymmetry is such that $D$, the dimensionality of the (Euclidean, in the present case) 
space-time of the Clifford algebra 
(\ref{weylclifford}) coincides with the
number $N$ of the extended supersymmetries, according to 
\begin{eqnarray}\label{weylcliffNsusycorr}
&\begin{tabular}{|c|c|c|} \hline
$\sharp$ of space-time dim. (Weyl-Clifford)&$\Leftrightarrow$& $\sharp$ of extended su.sies (in $1$-dim.)\\ \hline
$D$&=&$N$\\ \hline
\end{tabular}&\nonumber\\
\end{eqnarray}
The matrix size of the associated Clifford algebra (equal to $2d$, with $d$ given in (\ref{irrepdim})) 
corresponds to the number of (bosonic plus fermionic) fields entering the
one-dimensional $N$-extended supersymmetry irrep. \par
The classification of Weyl-type Clifford irreps, furnished in \cite{pt}, can be easily 
recovered from the well-known classification of Clifford irreps, given in \cite{abs} 
(see also \cite{por} and \cite{oku}).\par
The (\ref{length2irrep}) $Q_i$'s matrices realizing the $N$-extended supersymmetry algebra (\ref{susyalg}) 
on length-$2$ irreps have entries which are either $c$-numbers or are proportional to the hamiltonian $H$. 
Irreducible representations of higher length ($l\geq 3$) are systematically produced \cite{pt} through 
repeated applications  of the dressing transformations
\begin{eqnarray}\label{dressing}
Q_i &\mapsto & {{\widehat Q}_i}^{(k)} = S^{(k)}Q_i {S^{(k)}}^{-1}
\end{eqnarray}
realized by diagonal matrices $S^{(k)}$'s ($k=1,\ldots, 2d$) with entries ${s^{(k)}}_{ij}$
given by
\begin{eqnarray}\label{entries}
{s^{(k)}}_{ij} &=& \delta_{ij}(1-\delta_{jk}+\delta_{jk}H)
\end{eqnarray}
Some remarks are in order \cite{pt}\par
{\em i}) the dressed supersymmetry operators ${Q_i}'$ (for a given set of dressing transformations) 
have entries which are integral powers of $H$. A subclass of the ${Q_i}'$ s dressed operators 
is given by the local dressed operators,
whose entries are {\em non-negative} integral powers of $H$ (their entries have no $\frac{1}{H}$ poles). 
A local representation 
(irreps fall into this class) of an extended supersymmetry is realized by local dressed operators. 
The number of the extension, given by ${N}'$ (${N}'\leq N$), corresponds to the number 
of local dressed operators.\par
{\em ii}) The local dressed representation is not necessarily an irrep. Since the total number of fields 
($d$ bosons and $d$
fermions) is unchanged under dressing, the local dressed representation is an irrep iff $d$ and 
$N'$ satisfy the (\ref{irrepdim}) requirement (with $N'$ in place of $N$).\par
{\em iii}) The dressing changes the dimension (spin) of the fields of the original multiplet $m$. 
Under the $S^{(k)}$ dressing
transformation (\ref{dressing}), $m\mapsto S^{(k)}m$, all fields entering $m$ are 
unchanged apart the $k$-th one
(denoted, e.g., as $\varphi_k$ and mapped to ${\dot{\varphi_k}}$). Its dimension is 
changed from $[k]\mapsto [k]+1$.
This is why the dressing changes the length of a multiplet. As an example, 
if the original length-$2$ multiplet $m$ is a bosonic multiplet with $d$ spin-$0$ 
bosonic fields and $d$ spin-$\frac{1}{2}$ fermionic fields 
(in the following such a multiplet will be denoted as 
$(x_i;\psi_j)\equiv (d,d)_{s=0}$, for $i,j=1,\ldots, d$), then $S^{(k)}m$,
for $k\leq d$, corresponds to a length-$3$
multiplet with $d-1$ bosonic spin-$0$ fields, $d$ spin-$\frac{1}{2}$ fermionic fields and a single
spin-$1$ bosonic field (in the following we employ the notation $(d-1,d,1)_{s=0}$ for such a multiplet).\par
Let us fix now the overall conventions.
The most general multiplet is of the form $(d_1,d_2,\ldots , d_l)$, where $d_i$ for $i=1,2,\ldots ,l$
specify the number of fields of a given spin $s+\frac{i-1}{2}$. The spin $s$, i.e. the spin of the lowest
component fields in the multiplet, will also be referred to as the ``spin of the multiplet".
When looking purely at the representation properties of a given multiplet the assignment 
of an overall spin $s$ is arbitrary, since the supersymmetry transformations of the fields are 
not affected by $s$. Introducing a spin is useful for tensoring multiplets and becomes essential 
for physical applications, e.g. in the construction of supersymmetric invariant terms
entering an action.\par
In the above multiplet $l$ denotes its length, $d_l$ the number of auxiliary fields of 
highest spins transforming as time-derivatives. The total number of odd-indiced equal 
the total number of even-indiced fields, i.e.
$d_1+d_3+\ldots = d_2+d_4+\ldots = d$. The multiplet is bosonic 
if the odd-indiced fields are bosonic and the even-indiced are fermionic 
(the multiplet is fermionic in the converse case).
For a bosonic multiplet the auxiliary fields are bosonic (fermionic) if the length $l$ is an odd (even) 
number.\par
Just like the overall spin assignment, the assignment of a bosonic (fermionic) character to a multiplet
is arbitrary since the mutual transformation properties of the fields inside a multiplet are not affected by its
statistics. Therefore, multiplets always appear in dually related pairs s.t. to any bosonic multiplet there exists 
its fermionic counterpart with the same transformation properties (see also \cite{kra}).
\par
Throughout this paper we assign integer valued spins to bosonic multiplets 
and half-integer valued spins to fermionic multiplets. \par
As recalled before, the most general $(d_1,d_2,\ldots, d_l)$ multiplet is recovered 
as a dressing of its corresponding
$N$-extended length-$2$ $(d,d)$ multiplet. In \cite{pt} it was shown that all dressed supersymmetry 
operators producing any length-$3$ multiplet
(of the form $(d-p,d,p)$ for $p=1,\ldots, d-1$) are of local type. Therefore, for length-$3$ multiplets, 
we have $N'=N$.
This implies, in particular, that the $(d-p,d,p)$ multiplets are inequivalent irreps of the $N$-extended 
one-dimensional
supersymmetry. For what concerns length $l\geq 4$ multiplets, 
the general problem of finding irreps was not addressed in \cite{pt}.
It was shown, as a specific example, that the dressing of the length-$2$ $(4,4)$ irrep of $N=4$, 
realized through
the series of mappings $(4,4)\mapsto (1,4,3)\mapsto (1,3,3,1)$, produces at the end a length-$4$ 
multiplet $(1,3,3,1)$ 
carrying only three
local supersymmetries ($N'=3$). Since the relation (\ref{irrepdim}) 
is satisfied when setting equal to three the number
of extended supersymmetries and equal to $4$ the total number of bosonic 
(fermionic) fields entering a multiplet,
as a consequence, 
the $(1,3,3,1)$ multiplet corresponds to an irreducible representation of the $N=3$ extended supersymmetry.\par
In the next two sections, based on an algorithmic construction of representatives of Clifford
irreps, we present an iterative method to classify all irreducible representations of higher length
for arbitrary $N$ values of the extended supersymmetry (the complete results 
up to $N\leq 10$, plus some further selected cases, are explicitly presented in the appendices {\bf A} and {\bf B}).    
 
\setcounter{equation}{0}
\section{Oxidized supersymmetries}

In order to proceed at the classification of the length $l>3$ irreducible multiplets and attack the problem of
classifying the one-dimensional $N$ extended supersymmetries irreps, we need to use specific 
properties
of the associated Clifford irreps.\par 
We report here the needed mathematical background. We recall at first, see \cite{oku},
that the Clifford algebras generated by the $\Gamma$-matrices $\Gamma_i$, $i=1,\ldots, p+q$, satisfying
$\{\Gamma_i,\Gamma_j \}= 2\eta_{ij}$  for a $(p,q)$ signature (s.t. the $\eta_{ij}$ matrix is diagonal with
$p$ positive, $+1$, and $q$ negative, $-1$, entries), can be classified according to the most general matrix
$S$ commuting with all $\Gamma$'s (i.e. $[S,\Gamma_i]=0$ for any $i$). If the most general $S$ is a multiple of the
identity we get the normal (${\bf R}$) case. This situation occurs for $p-q =0,1,2~ mod ~8$. Otherwise,
for $p-q=3,7~ mod~ 8$, the most general $S$ is a sum of two matrices, the second one is a multiple of the
square root of $-1$ (this case is named the ``almost complex", {\bf C}, case) and, finally, for
$p-q= 4,5,6 ~ mod ~ 8$, the most general $S$ is a linear combination of four matrices closing the
quaternionic algebra (this case is referred to as the quaternionic, {\bf H}, case)\footnote{Throughout this paper we work 
with irreducible representations realized as real matrices. The ${\bf R}$, ${\bf C}$ and ${\bf H}$ cases, however, can 
also be described by matrices whose entries are valued in the corresponding division algebra.}.
\par
A real irreducible representation of the Clifford algebra is always unique \cite{oku} unless the relation
\begin{eqnarray}\label{irrcondition}
p-q&=& 1,5\quad mod\quad 8
\end{eqnarray}
is verified. For the above space-time signatures two inequivalent irreducible real representations are present,
the second one recovered by flipping the sign of all $\Gamma$'s ($\Gamma_i\mapsto -\Gamma_i$ for any $i$).\par
In the following, the Clifford irreps corresponding to the $(p,q)$ signatures are denoted as $Cl(p,q)$
(for our purposes there is no need to discriminate the two inequivalent irreps related with the (\ref{irrcondition})
signatures).\par
A concept that will be applied later is that of {\em maximal Clifford algebra} \cite{kt}. It corresponds to the
maximal number of Gamma matrices of $(p,q)$ signature which can be accommodated in a Clifford irrep of
a given matrix size. Non-maximal Clifford irreps are recovered from the maximal ones, after deleting a certain number
of Clifford Gamma matrices thought as external generators (see \cite{kt} for details). Maximal Clifford algebras can
also be referred to as the {\em oxidized} forms of a Clifford algebra, using a pun introduced in the superstrings/$M$-theory
literature, where {\em oxidation} denotes the inverse operation w.r.t. the dimensional reduction
\cite{oxi}.\par
Some remarks are in order.\par
{\em i}) An {\em oxidized} form of a Clifford algebra is encountered if and only if the
associated signature satisfies
the (\ref{irrcondition}) $p-q=1,5~ mod ~ 8$ condition.  \par
{\em ii}) Oxidized Clifford irreps are not of Weyl-type (see the previous section discussion). 
It is indeed always present, among their Clifford generators, the block-diagonal space-like Gamma 
matrix\footnote{Space-like gamma matrices $\gamma$ are those whose square is the identity; conversely,
a time-like gamma matrix $\gamma$ is such that $\gamma^2=-{\bf 1}_{2d}$, where ${\bf 1}_{2d}$ is the $2d\times 2d$
identity operator.}
$\left( \begin{tabular}{cc} ${\bf 1}$& $0$\\
$0$& $-{\bf 1}$
\end{tabular}
\right)$ which, on the light of the (\ref{weylcliffNsusycorr}) {\em Weyl-Clifford }$\Leftrightarrow$
{\em $1$-dim. $N$-extended su.sies irreps} correspondence, plays the role of a fermion number operator.
All the remaining generators of the oxidized Clifford irreps can be assumed of block-antidiagonal (Weyl type) form.\par
We can define as {\em oxidized} $N$-extended supersymmetries the ones constructed in terms of the supersymmetry generators 
associated,
according to (\ref{length2irrep}) and (\ref{weylclifford}), with the whole set of block-antidiagonal space-like 
gamma
matrices of a corresponding oxidized Clifford irrep.\par
The concept of {\em reduced} $N$-extended one-dimensional supersymmetries can be introduced in full analogy with the
concept of non-maximal Clifford algebras. The reduced $N$-extended supersymmetries are such that their length-$2$ irreps {\em do not} accommodate the whole maximal
number of supersymmetry generators at disposal. Stated otherwise, a reduced extended supersymmetry is always obtained 
from an associated oxidized
$N$-extended supersymmetry after deleting a certain number of supersymmetry
generators. Please notice that the irreducibility requirement is ensured if ${\widetilde N}$ (where ${\widetilde N}<N$ is
the number of reduced supersymmetry generators picked up from the whole set of generators of the oxidized extended
supersymmetry) satisfies a constraint related with the (\ref{irrepdim}) condition.  \par
From the results of \cite{kt} we can construct a first table, expressing the oxidized (and respectively the reduced) 
$N$-extended one-dimensional supersymmetries in terms of their associated oxidized Clifford irreps. We get 
{{
{ {{\begin{eqnarray}&\label{oxidation1}
\begin{tabular}{|c|l|l|}\hline
oxidized Clifford irreps & oxidized su.sies & reduced su.sies\\ \hline
$Cl(2+8m,1)_{\bf R}  $&$N=1+8m$&$\quad -$\\ \hline
$Cl(3+8m,2)_{\bf R}  $&$N=2+8m$&$\quad -$\\ \hline
$Cl(4+8m,3)_{\bf R}  $&$N=3+8m^{(\ast)}$&$\quad -$\\ \hline
$Cl(5+8m,0)_{\bf H}   $&$N=4+8m$&$N-1=3+8m^{(\ast\ast)}$\\ \hline
$Cl(6+8m,1)_{\bf H}  $&$N=5+8m^{(\ast\ast)}$&$\quad -$\\ \hline
$Cl(9+8m,0)_{\bf R}  $&$N=8+8m$& $N-1=7+8m$\\ 
&&$N-2=6+8m$\\
&&$N-3=5+8m^{(\ast)}$\\
\hline
\end{tabular}&\nonumber\\
&&\end{eqnarray}}} }  
}}
In the above table $m=0,1,2,\ldots$ is a non-negative integer. The oxidized Clifford irreps are of
real, ${\bf R}$, type (and, respectively, quaternionic, ${\bf H}$, type) if $p-q=1~ mod~ 8$ 
($p-q=5~ mod~ 8$).\par
It is worth mentioning that the oxidized and reduced extended supersymmetries are not affected by which one
of the two inequivalent choices for the (\ref{irrcondition}) Clifford irreps is made. As a consequence,
the length-$2$ irreps of the $N=1,2,4,6,7,8~mod~ 8$ extended supersymmetries are unique.
With respect to the $N=3,5 ~mod~8$ extended supersymmetries the situation is as follows.
For each such value of $N$ two inequivalent irreps are found (labeled, respectively, as $N=3^{(\ast)}, 3^{(\ast\ast)}~ mod~ 8$ and $N=5^{(\ast)}, 5^{(\ast\ast)}~ mod~ 8$) acting on multiplets with the same number of bosonic/fermionic fields.
The first class of ``$^{(\ast)}$" irreps corresponds to real-type supersymmetries, the second one (the ``$^{(\ast\ast)}$"
irreps) to quaternionic-type
supersymmetries. \par
Oxidized extended supersymmetries are found in the $N= 1,2,3^{(\ast)},4,5^{(\ast\ast)}, 8~ mod ~8$ cases.
Reduced extended supersymmetries are found for $N=3^{(\ast\ast)}~ mod~ 8$ (obtained from the $N=4
~mod~ 8$ oxidized form) and $N=5^{(\ast)}, 6,7~ mod~ 8$ (these three cases are recovered from the $N=8~mod~ 8$
class of oxidized extended supersymmetries). The whole picture is summarized in the following two
tables       
{{
{ {{\begin{eqnarray}&\label{oxidation2}
\begin{tabular}{|ccl|}\hline
Clifford irreps & $\rightarrow$ & Extended su.sies (mod 8)\\ \hline
$Cl(2+8m,1)_{\bf R}  $ & $\rightarrow$    &$N=1$\\ \hline
$Cl(3+8m,2)_{\bf R}  $ & $\rightarrow$    &$N=2$\\ \hline
$Cl(4+8m,3)_{\bf R}  $ & $\rightarrow$   &$N=3^{(\ast)}$\\ \hline
$Cl(5+8m,0)_{\bf H}  $  & $\rightarrow$   &$N=3^{(\ast\ast)}, 4$\\ \hline
$Cl(6+8m,1)_{\bf H}  $  & $\rightarrow$   &$N=5^{(\ast\ast)}$\\ \hline
$Cl(9+8m,0)_{\bf R}  $  & $\rightarrow$   &$N=5^{(\ast)},6,7,8$\\ 
\hline
\end{tabular}&\nonumber\\
&&\end{eqnarray}}} }  
}}
and 
{{
{ {{\begin{eqnarray}&\label{N3andN5}
\begin{tabular}{|lc|c|c|}\hline
$N=3^{(\ast)}$& mod $8$   & real    &oxidized\\ \hline
$N=3^{(\ast\ast)}$& mod $8$   & quaternionic    &reduced\\ \hline
$N=5^{(\ast)}$& mod $8$   & real    &reduced\\ \hline
$N=5^{(\ast\ast)}$& mod $8$   & quaternionic    &oxidized\\ \hline
\end{tabular}&\nonumber\\
&&\end{eqnarray}}} }  
}}
The fundamental $N=3^{(*)}, 3^{(\ast\ast)}, 5^{(\ast)},5^{(\ast\ast)}$ extended supersymmetries, whose complete list of
irreps is explicitly presented in Appendix {\bf A}, are obtained through\par
$Cl(4,3)\longrightarrow N=3^{(\ast)}$,\par
$Cl(5,0)\longrightarrow N=4\longrightarrow N=3^{(\ast\ast)}$,\par
$Cl(9,0)\longrightarrow N=8\longrightarrow N=5^{(\ast)}$,\par
$Cl(6,1)\longrightarrow N=5^{(\ast\ast)}$.\par
It is worth noticing that the $N=3^{(\ast)}$ oxidized supersymmetry (not admitting another Euclidean
supersymmetry generator) can however be extended to an oxidized pseudo-Euclidean generalized supersymmetry
(confront the discussion in Section {\bf 2}) with maximal number of six (with $(3,3)$ signature) pseudo-Euclidean
supersymmetry generators.\par
We recall, finally, that the dimensionality of the irreps of the $N$-extended supersymmetries can be read, for any $N$,
from (\ref{irrepdim}) (see also, for $N\leq 32$, the (\ref{N32}) table).\par

We conclude this section with some necessary remarks on the nature of the Clifford irreps.  A convenient way 
of systematically constructing a representative of each class of $Cl(p,q)$ irreducible Clifford representations
for any $(p,q)$ signature is through the algorithmic procedures, see \cite{crt2},
\par
{\em i}) 
$\gamma_i \mapsto \Gamma_j \equiv \left( \begin{tabular}{cc} $0$& $\gamma_i$\\
${\gamma}_i$& $0$
\end{tabular}
\right), \left( \begin{tabular}{cc} $0$& ${\bf 1}$\\
$-{\bf 1}$& $0$
\end{tabular}
\right),\left( \begin{tabular}{cc} ${\bf 1}$& $0$\\
$0$& $-{\bf 1}$
\end{tabular}
\right),$
mapping a $(p,q)$ Clifford irrep spanned by the $p+q$ $\gamma_i$'s matrices ($i=1,\ldots,p+q$) 
into a $(p+1,q+1)$ Clifford irrep and\par
{\em ii}) $\gamma_i \mapsto \Gamma_j \equiv \left( \begin{tabular}{cc} $0$& $\gamma_i$\\
$-{\gamma}_i$& $0$
\end{tabular}
\right), \left( \begin{tabular}{cc} $0$& ${\bf 1}$\\
${\bf 1}$& $0$
\end{tabular}
\right),\left( \begin{tabular}{cc} ${\bf 1}$& $0$\\
$0$& $-{\bf 1}$
\end{tabular}
\right),$ mapping the $(p,q)$ Clifford irrep into a $(q+2,p)$ Clifford irrep.\par
With the help of the two algorithms above, applied to the set of fundamental Clifford irreps
$1\equiv Cl(1,0)$, $Cl(0,3+8m)$ and $Cl(0,7+8m)$ (for $m=0,1,2,\ldots$) we can construct \cite{crt2} a representative
of any Clifford irrep for arbitrary values of $p$ and $q$. 
The set of $Cl(0,3+8m)$ and $Cl(0,7+8m)$ Clifford irreps were explicitly presented
in \cite{crt2}
as repeated tensor products of the set of the three real
$2\times 2$ matrices $\tau_1,\tau_2$ and $\tau_A$, given by
\begin{eqnarray}
&\tau_1=\left( \begin{tabular}{cc} $0$& ${1}$\\
${1}$& $0$
\end{tabular}
\right), \quad
\tau_2=\left( \begin{tabular}{cc} $1$& ${0}$\\
${0}$& $-1$
\end{tabular}
\right), \quad
\tau_A=\left( \begin{tabular}{cc} $0$& ${1}$\\
${-1}$& $0$
\end{tabular}
\right),&\nonumber
\end{eqnarray}
plus the $2\times 2$ identity ${\bf 1}_2$.\par
$Cl(0,3)$ is given by the three matrices $\tau_1\otimes \tau_A$, $\tau_2\otimes\tau_A$, $\tau_A\otimes{\bf 1}_2$,
while $Cl(0,7)$ is realized through the seven matrices  $\tau_1\otimes \tau_A\otimes {\bf 1}_2$,
$\tau_2\otimes \tau_A\otimes {\bf 1}_2$, 
$\tau_A\otimes {\bf 1}_2\otimes {\tau_1}$, $\tau_A\otimes {\bf 1}_2\otimes {\tau_2}$,
${\bf 1}_2\otimes \tau_1\otimes {\tau_A}$, ${\bf 1}_2\otimes \tau_2\otimes {\tau}_A$
and $\tau_A\otimes \tau_A\otimes {\tau}_A$.\par
$Cl(0,3+8m)$ (and, respectively, $Cl(0,7+8m)$), for $m=1,2,\ldots$, are recursively obtained as repeated
tensor products of $Cl(0,3)$ ($Cl(0,7)$) with $m$ sets of matrices from $Cl(1,8)$ and the ${\bf 1}_{16}$ identity
(see \cite{crt2} for details).\par
For our purposes it is sufficient to recall here that, as a consequence of the above algorithmic constructions,
the $Cl(p,q)$ real Clifford irreps present gamma matrices admitting one and only one non-vanishing entry
(given by $\pm 1$) in each column and in each row. Moreover, space-like gamma matrices do not
share the same non-vanishing entries; stated otherwise, there is no overlap,
an $e_{ij}\neq 0$ entry belongs to one, and only one, space-like gamma matrix.  
For any oxidized Clifford irrep we can further
compute the associated
``block-symbol", introduced in the next section, which allows us to systematically and efficiently count the
the number $N'$ of local supersymmetry dressed operators (see (\ref{dressing}) and the discussion thereafter).

\setcounter{equation}{0}
\section{Classification of the irreps}

In this section we present a systematic procedure to produce and classify length $l>3$ irreps of the (\ref{susyalg}) supersymmetry algebra for arbitrary values of $N$. We apply it to fully classify all irreps up to $N\leq 10$ and,
for the next cases of the oxidized $N=11^{(\ast)}$ and $N=12$ supersymmetries, the length-$4$ irreps (the results
are explicitly presented in the Appendices {\bf A} and {\bf B}). \par
Our approach is based on the following points (names and conventions here employed have been introduced in the
previous two sections):\par
{\em i}) the (\ref{length2irrep}) and (\ref{weylclifford}) connection between oxidized Clifford irreps and (oxidized and reduced)
length-$2$ irreps of the (\ref{susyalg}) supersymmetry algebra,\par
{\em ii}) the (\ref{dressing}) dressing transformation of length-$2$ irreps, producing length $l>2$ local type representations
of the (\ref{susyalg}) supersymmetry algebra,\par
{\em iii}) the matching condition (\ref{irrepdim}) between the number of the extended supersymmetries and the dimension of
the representation. It is satisfied if and only if the representation is irreducible and, finally,\par
{\em iv}) the algorithmic properties of the real Clifford irreps discussed at the end of Section {\bf 3}.\par
As explained in Section {\bf 2}, the dressing can produce $\frac{1}{H}$ poles in the dressed
supersymmetry operators. An $S^{(k)}$ dressing (\ref{dressing}, \ref{entries}) of a given supersymmetry operator $Q$
has the total effect of multiplying by $\frac{1}{H}$ 
all $Q$'s entries belonging to the $k$-th column and by $H$ all $Q$'s entries belonging to the $k$-th row, leaving unchanged
all remaining entries. In order to count (and remove) dressed operators with $\frac{1}{H}$ poles
one has to determine how non-vanishing entries are distributed in the whole set of supersymmetry operators
(since the $Q$'s are $2\times 2$ block-antidiagonal matrices, we can focus on the upper-right block,
the lower-left block presenting the same structure).
Up to $N\leq 8$, all non-vanishing entries of an oxidized supersymmetry fill the whole upper-right block
(for $N=8$, e.g., we have eight supersymmetry operators with $8$ non-overlapping non-vanishing entries each, s.t. $8\times8=64$,
filling the $8\times 8$ upper block chessboard of the $N=8$ supersymmetry).   
Starting from $N\geq 9$ this is no longer the case. The  $16\times 16$ right upper block ``chessboard" of the $N=9$
supersymmetry is filled with a total number of $9\times 16=144<16^2$ non-overlapping non-vanishing entries.\par
In the $N=9$ example
each column and each row of the upper-right (bottom-left) block intercepts the same amount of $9$ non-vanishing
entries belonging to the whole set of $9$ gamma matrices; the remaining $16-9=7$ entries are zero.\par
Not only the total number, but also the distribution of the non vanishing-entries inside the block matrices matters
when computing the locality condition of the dressed supersymmetry operators. The structure of the non-vanishing entries filling
the large-$N$ oxidized supersymmetries can be recovered from the algorithmic construction of the Clifford irreps
discussed in Section {\bf 3}. For $N\geq 8$, the filling of the upper-right block can be symbolically presented
(the block-symbol diagrams below)
in terms of the three fundamental fillings of an $8\times 8$ matrix. The three fundamental fillings, denoted as O, I, X,
represent, respectively,\par
\begin{tabular}{llll}
{\em i}) &O &$\equiv$& only vanishing entries,\\
{\em ii}) &I &$\equiv$ &non-vanishing entries filling the diagonal,\\
{\em iii}) &X &$\equiv$& non-vanishing entries filling the whole $8\times8$ matrix.
\end{tabular} 
\par
The block-symbols, explicitly presented here for the oxidized supersymmetries with $8\leq N\le 12$, are given by
\begin{eqnarray}
&\begin{tabular}{ccc}
$N=8$ &:&  $\left( \begin{tabular}{c} X \end{tabular}  \right)$\\
&&\\
$N=9$ &:& $\left(
\begin{tabular}{cc}
I&X\\
X&I
\end{tabular}
 \right)$\\
 &&\\
$N=10$ &:& $\left(
\begin{tabular}{cccc}
I&O&I&X\\
O&I&X&I\\
I&X&I&O\\
X&I&O&I
\end{tabular}
 \right)$\\
 &&\\
 $N=11^{\ast}$ &:& $\left(
\begin{tabular}{cccccccc}
I&O&O&O&I&O&I&X\\
O&I&O&O&O&I&X&I\\
O&O&I&O&I&X&I&O\\
O&O&O&I&X&I&O&I\\
I&O&I&X&I&O&O&O\\
O&I&X&I&O&I&O&O\\
I&X&I&O&O&O&I&O\\
X&I&O&I&O&O&O&I
\end{tabular}
\right)$\\
&&\\
$N=12$ &:& $\left(
\begin{tabular}{cccccccc}
I&X&I&O&I&O&I&O\\
X&I&O&I&O&I&O&I\\
I&O&I&X&I&O&I&O\\
O&I&X&I&O&I&O&I\\
I&O&I&O&I&X&I&O\\
O&I&O&I&X&I&O&I\\
I&O&I&O&I&O&I&X\\
O&I&O&I&O&I&X&I
\end{tabular}
 \right)$\\
\end{tabular}&
\end{eqnarray}
Block-symbols can be straightforwardly computed for arbitrary large-$N$ values of the oxidized
supersymmetries.\par
For reduced supersymmetries extra holes appear in the block-symbols, corresponding to the non-vanishing
entries belonging to the $N-N'$ supersymmetry operators that have been ``removed" from the whole set of
oxidized operators in order to produce the reduced $N'$-extended supersymmetry.  \par
Concerning multiplets, it is worth reminding that the diagonal dressing operator 
\begin{eqnarray}\label{fbduality}
S&=&\left(\begin{tabular}{cc}$H\cdot {\bf 1}_d$&$0$\\
$0$&${\bf 1}_d$\end{tabular}
\right)
\end{eqnarray} 
applied on a $(d,d)$ length-$2$ multiplet reverses its statistics (the same transformation reverses the statistics 
of fields in any given multiplet).\par
Length-$3$ multiplets are obtained by applying, on a $(d,d)$ length-$2$ multiplet, diagonal
dressing operators $S$ with a total number of $k$ (with $1\leq k\leq d-1$) single powers of $H$ in the first 
$d$ diagonal entries, while
the $2d-k$ remaining diagonal entries are $1$.\par
Length-$4$ multiplets require dressing operators with ${\tilde k}$ (for $1\leq{\tilde k}\leq d-1$)
single powers of $H$ diagonal entries in the positions $d+1,\ldots, 2d$.
\par
Length-$5$ (length-$6$) multiplets require a dressing operator $S$ with at least one $H^2$ second power
diagonal entry in the position $1,\ldots,d$ (and, respectively, $d+1,\ldots, 2d$). \par
Length-$7$ and length-$8$ multiplets require dressing operators with at least a third power, $H^3$, 
diagonal entry and so on. \par
We are now in the position to compute the length $l\geq 4$ irreducible representations of the
oxidized supersymmetries. Let us illustrate at first an $N=9$ example.
An $N=9$ length-$3$ irrep with $15$ auxiliary fields (i.e. $(1,16,15)$) is such that the original
$(16,16)$ upper-right
block ${\cal B}$ is mapped into a new block, ${\cal B}\mapsto{\cal B}'$, by multiplying $15$ columns by $H$,
while   
leaving the remaining column unchanged. The lengthening $3\mapsto 4$,
obtained by leaving unchanged the number of fields, $15$, in the third position, produces a block-mapping ${\cal B}' \mapsto
{\cal B}''$, where the new block is obtained from ${\cal B}'$ by multiplying a certain number of rows by $\frac{1}{H}$,
while the remaining ones are left unchanged. The condition that no ${\frac{1}{H}}$ poles appear in ${\cal B}''$
implies that, at most,
seven rows can be picked up. They have to be chosen among the ones corresponding to the zeroes of the single, unchanged, column
of ${\cal B}'$. It turns out that $N=9$ admits seven inequivalent length-$4$ irreps of the type
$(1,16-k,15,k)$, for $k=1,2,\ldots , 7$. \par
The same strategy can be applied starting from $(2,16,14)$, $(3,16,13)$ and so on. At the end we produce the complete
list of length-$4$ irreps of $N=9$ (listed in Appendix {\bf B}). This procedure straightforwardly works for 
computing length-$4$ irreps of any oxidized value of $N$, once the corresponding block-symbols
are known.\par
For what concerns $l>4$, let us illustrate the $N=10$ length-$5$ example, since $10$ is the least value
of an extended supersymmetry admitting irreps with $l>4$.
Let us check, at first, whether we can produce a single auxiliary field in the fifth position. This amounts to multiply
by $\frac{1}{H^2}$ a single row of the original $(32,32)$ bottom-left block. Since all its entries, see (\ref{length2irrep}), are
already multiplied by $H$, this implies that the new bottom-left block admits a single $\frac{1}{H}$ pole in correspondence with the non-vanishing entries of the transformed row, while it is regular anywhere else.
We get on the transformed row ten poles.
In order to kill them we need to multiply (at least) the $10$ corresponding columns of the bottom-left
block by $H$. This multiplication corresponds to the transformation which maps (at least) $10$ fields from the
second to the fourth position. This transformation acts on the upper-right block by multiplying the corresponding rows
by $\frac{1}{H}$. In its turn, these extra-poles have to be cancelled by multiplying a convenient number of columns by $H$
(in correspondence with the transformation mapping fields from the first to the third position). The
extra $\frac{1}{H}$ poles produced by this new compensating transformation on the corresponding rows of the bottom-left 
block do not produce any further
singularity, due to the presence of the
overall $H$ factor mentioned above. \par
The same procedure can be later applied to verify whether there is enough room to have two, three or more fields in the fifth position. \par
Length $l\geq 6$ irreps can be analyzed along the same lines. The complete result for the $N=10$ irreps is furnished in
Appendix {\bf B}. \par
For what concerns the reduced extended supersymmetries, the computation of their irreps can be carried on just like the
oxidized supersymmetries, but taking into account that their block-symbols admit extra holes. We concentrate
on $N=8$ reductions. The eight gamma matrices generating $N=8$ under the (\ref{weylcliffNsusycorr}) correspondence are all
on equal footing. We can single out any one of them (let's say the one with a diagonal upper-right block) in order that
the remaining ones generate $N=7$. The diagonal holes in the $N=7$ block-symbol imply that, just like 
the first $N=9$ example discussed above, we can lengthen the $N=8$ $(1,8,7)$ irrep into an $N=7$ $(1,7,7,1)$
irrep. The analysis of the $N=5^{(\ast)},6$ (and $N=3^{(\ast\ast)}$ derived from $N=4$) cases is done in the
same way. 
\par
Let us now make some necessary remarks on the irreducible representations. 
Two types of dualities act on them.
We have at first the {\em fermion} $\Leftrightarrow$ {\em boson} duality, obtained by exchanging, via the
(\ref{fbduality}) dressing, the statistics of the component fields in the multiplet. A second type
of duality can be referred to as the {\em high} $\Leftrightarrow$ {\em low} spin duality.
This new duality involves the mapping of a $(d_1,d_2,\ldots ,d_l)$ irreducible multiplet into its
irreducible dual multiplet
\begin{eqnarray}\label{hiloduality}
(d_1,d_2,\ldots ,d_l) &\Leftrightarrow& (d_l,d_{l-1},\ldots ,d_1)
\end{eqnarray}
obtained by turning the highest-spin fields into the lowest spin fields. Therefore this duality relates two 
opposite statistics multiplets if $l$ is even and two multiplets with the same statistics if $l$ is odd.\par
Let us denote with $(~^{1}x_{j_1}; ~^2x_{j_2}; \ldots ; ~^{l}x_{j_l})$ the set of fields entering
$(d_1,d_2,\ldots ,d_l)$ (here $j_i=1,\ldots ,d_i$). The dual irreducible $(d_l,d_{l-1},\ldots ,d_1)$ multiplet can be realized
with the fields $(~^lx_{j_l};~ ^{l-1}{\dot x}_{j_{l - 1}}; \ldots ; ~^1{x_{j_1}}^{(l-1)})$, where $x^{(k)}$ here denotes
the application of the time derivative $k$-times. Applying the same transformation on the latter multiplet we obtain
a new multiplet, $({~^{1}x_{j_1}}^{(l-1)}; ~^2x_{j_2}^{(l-1)}; \ldots ; ~^{l}x_{j_l}^{(l-1)})$, whose supersymmetry transformations are nevertheless the same as the original ones. As a corollary, the class of the irreducible representations
is closed under the (\ref{hiloduality}) {\em high $\Leftrightarrow$ low} spin duality.
\par
The {\em high $\Leftrightarrow$ low spin} duality (\ref{hiloduality}) concides with the {\em fermion $\Leftrightarrow $ boson}
(\ref{fbduality}) duality only when applied to self-dual (under (\ref{hiloduality})) multiplets of even length. It is a distinct 
duality transformation in the remaining cases.

\par
For what concerns the total number ${\overline{\kappa}}$ of inequivalent irreps of the $N$-extended supersymmetry, it
is given by the sum of the ${\overline{\kappa}}_l$ inequivalent irreps of length-$l$, namely, 
\begin{eqnarray}{\label{kappa}}
{{\overline\kappa}} &=& \sum_{l=2}^L {{\overline\kappa}}_l
\end{eqnarray}
where $L$ is the maximal length for an $N$-extended supersymmetry irrep.\par
${\overline k}$ is the counting of inequivalent irreps irrispectively of the overall statistics of the multiplets.
A factor $2$ can be introduced if we want to discriminate
the statistics of the multiplets (bosonic or fermionic). In this case
the number of inequivalent irreps is $\kappa$, with
\begin{eqnarray}\label{ferbosin}
{\kappa}&=& 2{\overline \kappa}
\end{eqnarray} 
Let us present now a series of results concerning the irreducible irreps
(a more detailed list can be found in the Appendices {\bf A} and {\bf B}).\par
Up to $N\leq 8$, length-$4$ irreps are present only for reduced supersymmetries. The complete
list of length-$4$ irreps up to $N=8$ is given by 
{{
{ {{\begin{eqnarray}&\label{N8length4}
\begin{tabular}{|l|c|}\hline
 $N=1$&NO\\ \hline
 $N=2$&NO\\ \hline
$N=3   $&$(1,3,3,1)$\\ \hline
$N=4$&NO\\ \hline
$N=5   $&$(1,5,7,3), ~(3,7,5,1),~(1,6,7,2), ~(2,7,6,1),~(2,6,6,2), ~(1,7,7,1)$\\ \hline
$N=6   $&$(1,6,7,2),~(2,7,6,1),~(2,6,6,2), ~(1,7,7,1)$\\ \hline
$N=7   $&$(1,7,7,1)$\\  \hline
$N=8$&NO\\ \hline
\end{tabular}&\nonumber\\
&&\end{eqnarray}}} }  
}}
Since there are no length-$l$ irreps with $l\geq 5$ for $N\leq 9$, the above list, together
with the already known length-$2$ and length-$3$ irreps, provides the complete classification
of inequivalent irreps for $N\leq 8$. \par
Please notice that the length-$4$ irrep of $N=3$, $(1,3,3,1)$, is self-dual under the
(\ref{hiloduality}) {\em high} $\Leftrightarrow$ {\em low} spin duality, while two of the inequivalent 
length-$4$ $N=5$ irreps are self-dual, $(2,6,6,2)$ and $(1,7,7,1)$. The remaining ones
are pair-wise dually related ($(1,5,7,3)\Leftrightarrow (3,7,5,1)$ and $(1,6,7,2)\Leftrightarrow (2,7,6,1)$).\par
The list of inequivalent length-$4$ irreps is the same for both derivations (real and quaternionic,
see the previous section discussion) of the $N=3$ and $N=5$ extended supersymmetries. It is however convenient
to distinguish between real and quaternionic derivations of the $N=3,5\quad mod\quad 8$ extended supersymmetries,
due to their different properties. As an example, the $(1,3,3,1)$ length-$4$ irrep of the
$N=3^{(\ast)}$ supersymmetry can be oxidized to a length-$4$ irrep of the $(3,3)$ pseudosupersymmetry (\ref{susyalg}),
while the corresponding quaternionic $(1,3,3,1)$ $N=3^{(\ast\ast)}$ irrep cannot be oxidized to a pseudosupersymmetry.
Similarly, the quaternionically derived length-$4$ irreps of the $N=5^{(\ast\ast)}$ supersymmetry are oxidized
to length-$4$ irreps of the $(5,1)$ extended pseudosupersymmetry. For what concerns the real length-$4$ irreps of
the $N=5^{(\ast)}$ supersymmetry the picture is the following.
Due to the reduction chain from the $N=8$ oxidized supersimmetry 
\begin{eqnarray}
&N=8\rightarrow N=7
\rightarrow N=6\rightarrow N=5^{(\ast)}
&
\end{eqnarray}
it turns out that the $(1,7,7,1)$ irrep of $N=5^{(\ast)}$ can be {\em oxidized} as an $N=6$ and $N=7$ irrep.
The $(1,6,7,2)\Leftrightarrow (2,7,6,1)$ and $(2,6,6,2)$ multiplets, thought as $N=5^{(\ast)}$ irreps, can
be oxidized and promoted to be $N=6$ irreps.  \par
In the Appendix {\bf B} the complete classification
of inequivalent irreps for $N=9,10$ is presented. Therefore, we are able to produce here another table, expressing
the maximal length $L$ and the total number ${\overline \kappa}$ of inequivalent irreps for the $N$-extended
supersymmetries with $N\leq 10$. We have 
{{
{ {{\begin{eqnarray}&\label{inequivalent}
\begin{tabular}{|l|c|c|}\hline
su.sies& $L$&${\overline\kappa}_2+\ldots +{\overline\kappa}_L={\overline\kappa}$\\ \hline 
$N=1   $&$2$& $ 1$\\ \hline
$N=2   $&$3$&$1+1=2$\\ \hline
$N=3   $&$4$&$1+3+1=5$\\ \hline
$N=4   $&$3$&$ 1+3=4$\\ \hline
$N=5  $&$4$&$1+7+6=14$\\ \hline
$N=6  $&$4$&$1+7+4=12$\\ \hline
$N=7   $&$4$&$1+7+1=9$\\ \hline
$N=8  $&$3$&$1+7=8$\\ \hline
$N=9  $&$4$&$1+15+28=44$\\ \hline
$N=10  $&$5$&$1+31+176+140=348$\\ \hline
\end{tabular}&\nonumber\\
&&\end{eqnarray}}} }  
}}
We conclude this section pointing out that 
the procedure here outlined can be systematically carried on to fully classify inequivalent irreps
for arbitrarily large values of $N$; the limitations are only due to the increasing of the required computational work.
   
\setcounter{equation}{0}
\section{On the ``enveloping" representations}

For each $N$, we can introduce the ``enveloping representation" of the $N$-extended one-dimensional supersymmetry
as the representation given by $2^{N-1}$ bosonic and $2^{N-1}$
fermionic states spanned by the monomials
\begin{eqnarray}\label{monomials}
&\prod_{i=1}^N {Q_i}^{\alpha_i},&\nonumber
\end{eqnarray}
where the $\alpha_i$'s take the values $0$ and $1$.\par
The state with $\alpha_i=0$ for any $i$ is a bosonic state of spin $s=0$ and corresponds to the identity operator
${\bf 1}$. The $\left(
\begin{array}{c}
N
\\
k	
\end{array}
 \right)$
states given by $Q_{i_1}\cdot Q_{i_2}\cdot\ldots Q_{i_k}\cdot {\bf 1}$ (all $i_j$'s are different) belong to spin $s=\frac{k}{2}$.\par
For $N=1,2,3$, the enveloping representation is irreducible. It can be identified with the bosonic
irreducible multiplets $(1,1)$, $(1,2,1)$ and $(1,3,3,1)$, respectively. 
The last multiplet is the unique length-$4$ multiplet of $N=3$, see the previous section results.\par
Starting from
$N\geq4$, the enveloping representation is no longer irreducible. The $N=4$ enveloping representation
corresponds to the bosonic multiplet $(1,4,6,4,1)$, which admits the
following   
decomposition into its irreducible components
\begin{eqnarray}{\label{adjointdecomp}}
(1,4,6,4,1)_{s=0} &\equiv& (1,4,3)_{s=0} + (3,4,1)_{s=1}
\end{eqnarray}
The explicit decomposition of the $N=4$ enveloping representation into its irrep constituents
(also discussed in Appendix {\bf C}, see the {\em iia}) case) requires the knowledge of the
(\ref{N4constraints}) constraints below.\par
All $N=4$ irreps satisfy
\begin{eqnarray}{\label{N4constraints}}
Q_1Q_2 &=& Q_3Q_4\Gamma^5,\nonumber\\
Q_2Q_3 &=& Q_1Q_4\Gamma^5,\nonumber\\
Q_3Q_1 &=& Q_2Q_4 \Gamma^5,
\end{eqnarray}
where $\Gamma^5=\left(\begin{array}{cc} {\bf 1}_4& 0\\
0 &-{\bf 1}_4
\end{array}\right)$ plays the role of the fermion number operator
\begin{eqnarray}{\label{fermionnumber}}
\Gamma^5 (boson) &=& +(boson),\nonumber\\
\Gamma^5 (fermion) &=& -(fermion)
\end{eqnarray}
for any bosonic (fermionic) state in the multiplet.\par
The set of (\ref{N4constraints}) equations can be easily verified on the length $2$
$(4,4)$ irrep. It holds also for any length $3$ irrep, since in each case the dressing transformations
(\ref{dressing})
$Q_i \mapsto SQ_iS^{-1}$ discussed in Section {\bf 2} are realized by diagonal matrices
$S$ which commute with $\Gamma^5$.

\setcounter{equation}{0}
\section{Construction of invariants for potential and kinetic terms}

The knowledge of all finite, linear multiplets of irreducible representations of the $N$-extended supersymmetries
allows us to systematically construct invariants for any $N$-extended supersymmetry. In this section we point out how
this can be done. We recall that those component fields in 
any given multiplet corresponding to the highest spin can be regarded as auxiliary fields. They transform, under 
each one of the $N$ supersymmetries, as time-derivatives. They can therefore be picked up us lagrangian terms which, inside an
action, provide a manifest invariant for the $N$-extended supersymmetry.\par
The linearity of the supersymmetry transformations implies that the tensored multiplets, obtained as bilinear, trilinear or
in general $k$-linear products\footnote{We remark that, due to the renormalizability condition, the supersymmetric actions 
of the dimensional reduction to $D=1$ of the renormalizable four-dimensional supersymmetric field theories admit terms which are at most quartic in the fields. As for the total number of extended supersymmetries of the one-dimensional reduced models, it is four times the number of supersymmetries in $D=4$. Therefore, the maximal $N=4$ four-dimensional SuperYang-Mills theories are reduced to $N=16$
supersymmetric $1D$ systems, while the $1D$ dimensional reduction of the (non-renormalizable)
maximal $N=8$ supergravity gives an $N=32$ system.} of the original irreps, keep the same structure as the original irreps. In particular they
can be decomposed into their irreducible component multiplets, which provide the associated multilinear invariants.
Specific kinds of such invariants include the kinetic terms, as well as the potential terms (their multilinear invariants are 
functions of the original component
fields alone and do not involve their time derivatives). 
For illustrative purposes we present here the construction of the invariants in some selected examples. The procedure
here outlined can be carried out sistematically, without any conceptual problem, just increasing of computational
work, for arbitrarily large values of $N$.  The method here discussed allows to construct manifest invariants of the
$N$-extended supersymmetries, {\em without} introducing a superspace and a superfield formalism, recovering the same
results of the superspace construction for small values of $N$, but allowing to extend it when no
superspace formalism is available (for $N>8$, see \cite{abc}).         \par
Let us discuss at first the multi-linear invariants associated with the two inequivalent irreps of the $N=2$ supersymmetry
(see Appendix {\bf C}).
$k$ products of the same $(1,2,1)_{s=0} = (x;\psi_1,\psi_2; g)$ irrep produce the spin $s=0$ $k$-linear $(1,2,1)$
irrep given by
\begin{eqnarray}\label{klinear}
(x_k;{\psi_1}_k,{\psi_2}_k;g_k)&=& (x^k; k\psi_1x^{k-1},k\psi_2x^{k-1};kgx^{k-1}-k(k-1)x^{k-2}\psi_1\psi_2)
\end{eqnarray}
A single $N=2$ invariant is produced at spin $s=1$. It is given by
\begin{eqnarray}\label{invariants}
I &=& \int dt \left(\sum_{k=1}^{\infty} c_k g_k\right)
\end{eqnarray}
with arbitrary constants $c_k$'s. This $N=2$ invariant corresponds to the most general self-interacting potential of a
single $(1,2,1)$ real superfield. \par
Multilinear invariants for the spin $s=0$  $(2,2)_{s=0}\equiv(x_1,x_2;\psi_1,\psi_2)$ chiral superfield in our framework can be constructed as follows.
The unique bilinear invariant at $s=1$ for the $(2,2)_{s=0}$ irrep
is obtained (see Appendix {\bf C}, case $ic$)) from the spin $s=1$ term in the $(1,2,1)_{\parallel s=0}$
entering the r.h.s. decomposition of $(2,2)_{s=0}\times (2,2)_{s=0}$ after identifying the left and right tensored 
multiplets.
We get for the corresponding auxiliary field
\begin{eqnarray}\label{mixedaux}
{\hat g} &=& 2x_1{\dot x}_2-2x_2{\dot x}_1 + 4\psi_2\psi_1
\end{eqnarray}
Two spin $s=1$ trilinear invariants are obtained in two equivalent ways,  
either from the $(1,2,1)_{\parallel s=0}^{(3)}$ and $(1,2,1)_{\perp s=0}^{(3)}$
multiplets in the irrep decompositions of the tensor product
$(2,2)_{s=0}\times {(2,2)_{\parallel s=0}}^{(2)}$, 
or from the two ${{(1,2,1)_{s=0}}^{a,b }}^{(3)}$ multiplets in the irrep decompositions of the $(2,2)_{s=0}\times {(1,2,1)_{\parallel s=0}}^{(2)}$ product 
(here $(\ldots)^{(k)}$ specifies an irrep which is $k$-linear w.r.t. the original fields and 
${(2,2)_{\parallel s=0}}^{(2)}$, ${(1,2,1)_{\parallel s=0}}^{(2)}$ denote the corresponding irreps in the
$(2,2)_{s=0}\times (2,2)_{s=0}$ decomposition). 
\par
In both cases we get the auxiliary fields
\begin{eqnarray}\label{twoaux}
g_{I} &=& -8x_1\psi_1\psi_2 -2{\dot x}_1x_1x_2+{\dot x}_2({3x_1}^2+{x_2}^2)\nonumber\\
g_{II} &=& -8x_2\psi_1\psi_2 +2{\dot x}_2x_1x_2-{\dot x}_1({3x_2}^2+{x_1}^2)
\end{eqnarray}
Please notice that $g_I$, $g_{II}$ are mutually recovered by exchanging $x_1\leftrightarrow -x_2$, 
$\psi_1\leftrightarrow\psi_2$.\par
Spin $s=1$ $k$-linear self-invariants of the $N=2$ chiral superfield can be recursively constructed by
applying the same scheme. Please notice that such invariants do not fall into the class of potential invariants
since they involve time-derivatives of the original fields.\par
It is convenient to illustrate now the next simplest example of invariant, given by the self-interaction 
of the spin $s=0$ $(1,4,3)$ irrep of the $N=4$ extended supersymmetry. According to the $iia$) case of Appendix {\bf C}
the tensor product of two $(1,4,3)$ irreps (which, for our purposes here, are identified) gives rise to an enveloping 
representation of $N=4$.  It contains a single auxiliary field at $s=2$ which generates the $N=4$ invariant.
In our case the auxiliary field is given by
\begin{eqnarray}
K& = &-{\ddot x}x -{\psi_1}{\dot\psi}_1-\psi_2{\dot\psi}_2-\psi_3{\dot\psi}_3-\psi_4{\dot\psi}_4+
{g_1}^2+{g_2}^2+{g_3}^2
\end{eqnarray}
and coincides with the kinetic term for the $(1,4,3)$ irrep.\par
An alternative construction is however available, due to the fact that the enveloping $N=4$ representation is reducible and
can be decomposed into its irreps according to (\ref{adjointdecomp}).
The eight bilinear terms entering the $(3,4,1)_{s=1}$ multiplet in the
irrep decomposition of the $N=4$ enveloping representation can be consistently
set all equal to zero. The surviving bilinear terms enter, see (\ref{adjointdecomp}), the $(1,4,3)_{s=0}$ irrep
admitting three auxiliary fields (and, therefore, three associated invariants) at spin $s=1$.
In terms of {\em bilinearly constrained} $(1,4,3)$ component fields we obtain three invariants, associated with the   
three auxiliary fields $a_1$, $a_2$, $a_3$ given by
\begin{eqnarray}\label{123inv}
a_1 &=& \psi_2\psi_4 -g_2 x\nonumber\\
a_2 &=& \psi_2\psi_3 +g_3 x\nonumber\\
a_3 &=& \psi_2\psi_1 -g_1 x
\end{eqnarray}
together with eight bilinear constraints given by
\begin{eqnarray}\label{8constr}
&& C_1  = \psi_4\psi_2 +\psi_3\psi_1 =0\nonumber\\
&& C_2  = \psi_4\psi_3 +\psi_1\psi_2 =0\nonumber\\
&& C_3  = \psi_4\psi_1 +\psi_2\psi_3 =0\nonumber\\
&& C_4  = {\dot\psi}_1x -g_2\psi_3-g_3\psi_4-g_1\psi_2=0\nonumber\\
&& C_5  = {\dot\psi}_3x +g_2\psi_1-g_1\psi_4+g_3\psi_2=0\nonumber\\
&& C_6  = {\dot\psi}_4x +g_3\psi_1+g_1\psi_3-g_2\psi_2=0\nonumber\\
&& C_7  = {\dot\psi}_2x -g_1\psi_1+g_3\psi_3-g_2\psi_4=0\nonumber\\
&& C_8  = -{\ddot x}x -{\psi_1}{\dot\psi}_1-\psi_2{\dot\psi}_2-\psi_3{\dot\psi}_3-\psi_4{\dot\psi}_4+
{g_1}^2+{g_2}^2+{g_3}^2=0
\end{eqnarray}
This procedure can be straightforwardly iterated to produce three spin $s=1$ multilinear invariants
and eight multilinear constraints. \par
The first three constraints ($C_1,C_2,C_3$) contain bilinear terms in the Grassmann fields $\psi_i$'s. Since we cannot take a quotient of a Grassmann field, these constraints cannot be solved algebraically.
For what concerns the remaining constraints, we can use $C_4$ and $C_5$ to express, algebraically,
$\psi_2$ and $\psi_4$ in terms of the remaining fields. We get
\begin{eqnarray}\label{psi2and4}
\psi_2 &=& \frac{1}{{g_1}^2 + {g_3}^2}\left[g_1x{\dot \psi}_1-g_3 x{\dot \psi}_3 -g_1g_2\psi_3-g_2g_3\psi_1 \right]\nonumber\\
\psi_4 &=& \frac{1}{{g_1}^2 + {g_3}^2}\left[g_1x{\dot \psi}_3+g_3 x{\dot \psi}_1 +g_1g_2\psi_1-g_2g_3\psi_3 \right]
\end{eqnarray}
The resulting supersymmetry is realized non-linearly
in terms of four bosonic and two fermionic fields. The remaining constraints $C_i$, for $i=1,2,3,6,7,8$,
can no longer be solved algebraically since, expressing $\psi_2$ and $\psi_4$ with the (\ref{psi2and4})
relations, they necessarily involve time derivatives of all fields. 
It can be easily seen that, by taking the most general invariant action of the 
$(1,4,3)$ multiplet (see e.g. (\ref{N4geninvact})) and setting the bilinear constraints ({\ref{8constr}) to hold
at the initial time $t_0$, the constraints are preserved by the equations of motion at any later instant $t>t_0$.\par  
We can discuss along the same lines the multilinear invariants for the self-interacting $N=4$ spin $s=0$
$(2,4,2)_{s=0}\equiv (x_1,x_2;\psi_1,\psi_2,\psi_3,\psi_4;g_1,g_2)$ irrep.  Due to the $iic$) decomposition of Appendix {\bf C} we obtain, in terms of
{\em unconstrained} component fields, two bilinear invariants at spin $s=1$, associated with the auxiliary
fields
\begin{eqnarray}\label{s1twoinv}
{\tilde g}_{I}&=& g_1x_1-g_2x_2-\psi_2\psi_4+\psi_1\psi_3,\nonumber\\
{\tilde g}_{II} &=& g_2x_1+g_1x_2+\psi_1\psi_4+\psi_2\psi_3
\end{eqnarray}
plus the kinetic invariant (with ${\tilde K}$ as kinetic density) at spin $s=2$ arising from the enveloping representation
\begin{eqnarray}\label{kininv}
{\tilde K} &=& {\dot x}_1{\dot x}_1+{\dot x}_2{\dot x}_2 
-\psi_1{\dot\psi}_1-\psi_2{\dot\psi}_2-\psi_3{\dot\psi}_3-\psi_4{\dot\psi}_4
+{g_1}^2+{g_2}^2
\end{eqnarray}
As in the previous case, we can consistently introduce eight bilinear constraints
in relation with the decomposition into irreps of the $N=4$ enveloping representation.
As a result, three extra bilinear invariants for the bilinearly {\em constrained} fields 
are obtained at spin $s=1$. The three invariants are associated with the auxiliary fields
\begin{eqnarray}\label{s1threeinv}
{\tilde a}_1 &=& \psi_2\psi_4 -g_1x_1-g_2 x_2+\psi_1\psi_3\nonumber\\
{\tilde a}_2 &=& \psi_2\psi_3 ++g_2x_1-\psi_1\psi_4-g_1x_2\nonumber\\
{\tilde a}_3 &=& 2\psi_2\psi_1 -{\dot x}_2x_1+{\dot x}_1x_2
\end{eqnarray}
while the eight bilinear constraints are explicitly given by
\begin{eqnarray}\label{8newconstr}
&& {\widetilde C}_1  = \psi_2\psi_4+\psi_1\psi_3 =0\nonumber\\
&& {\widetilde C}_2  = \psi_4\psi_3+\psi_1\psi_2 =0\nonumber\\
&& {\widetilde C}_3  = \psi_1\psi_4+\psi_3\psi_2 =0\nonumber\\
&& {\widetilde C}_4  = x_1{\dot \psi}_1-{\dot x}_1\psi_1+x_2{\dot\psi}_2-{\dot x}_2\psi_2-2g_1\psi_3-2g_2\psi_4=0\nonumber\\
&& {\widetilde C}_5  = x_1{\dot \psi}_3-{\dot x}_1\psi_3+x_2{\dot\psi}_4-{\dot x}_2\psi_4+2g_1\psi_1+2g_2\psi_2 =0\nonumber\\
&& {\widetilde C}_6  = x_1{\dot \psi}_4-{\dot x}_1\psi_4-x_2{\dot\psi}_3+{\dot x}_2\psi_3-2g_1\psi_2+2g_2\psi_1 =0\nonumber\\
&& {\widetilde C}_7  = x_1{\dot \psi}_2-{\dot x}_1\psi_2-x_2{\dot\psi}_1+{\dot x}_2\psi_1+2g_1\psi_4-2g_2\psi_3 =0\nonumber\\
&& {\widetilde C}_8  = {{\dot x}_1}^2+{{\dot x}_2}^2 -{\ddot x}_1x_1-{\ddot x}_2x_2 -2\psi_1{\dot \psi}_1
-2\psi_2{\dot\psi}_2-2\psi_3{\dot\psi}_3-2\psi_4{\dot\psi}_4+2{g_1}^2+2{g_2}^2=0\nonumber
\end{eqnarray}
\begin{eqnarray}
&&
\end{eqnarray}
Just like in the self-interacting $(1,4,3)$ case, the last bilinear constraint coincides with the kinetic
density.\par
Summarizing, either we work with unconstrained fields and obtain two spin $s=1$ invariants
plus an invariant kinetic term, or we work with constrained fields, obtaining $5=2+3$ invariants at
spin $s=1$ associated with $8$ bilinear constraints.\par
Let us discuss now a general construction of the (constant) invariant kinetic term for arbitrary $N$
(the cases previously discussed were specific of the $N=4$ case, since ``accidentally" the $N=4$ enveloping
representation admits a spin $s=2$ auxiliary field).\par
Let us consider a generic length $l\leq 4$ spin $s=0$ irrep of the $N$-extended supersymmetry, given by (see Appendix {\bf A})
$(d-p,d-q,p,q) \equiv (x_i;\psi_j;g_k;\omega_l)$, where the $x$'s and $g$'s are bosonic spin $s=0$ and respectively
$s=1$ component fields, while the $\psi$'s and $\omega$'s are fermionic spin $s=\frac{1}{2}$ (respectively $s=\frac{3}{2}$)
component fields (for length $2$ and $3$ irreps there are no $\omega$'s fields, namely $q=0$).\par
The kinetic density has dimension $2$ (we recall, see the discussion in Section {\bf 2}, that we use,
interchangeably, the words ``spin" and ``dimension"). It can be symbolically written, dropping the field indices, as ${\dot x}^2+\psi{\dot\psi}+g^2+\omega\psi$.
In order to produce a spin $s=2$ auxiliary field which can be interpreted as a kinetic density we proceed as follows.
At first we transform the $(x_i;\psi_j;g_k;\omega_l)_{s=0}$ irrep into \par
{\em i}) a bosonic spin $s=1$ length-$2$ irrep $(d,d)_{B, s=1}$, through the mapping\par
$(x_i;\psi_j;g_k;\omega_l)\mapsto ({\dot x}_i, g_k; {\dot \psi}_j, \omega_l)\equiv (d,d)_{B, s=1}$ and\par
{\em ii}) a fermionic spin $s=\frac{1}{2}$ length-$3$ irrep $(d-q,d, q)_{F, s=\frac{1}{2}}$, through the mapping\par
$(x_i;\psi_j;g_k;\omega_l)\mapsto (\psi_j; {\dot x}_i, g_k; \omega_l)\equiv (d-q,d, q)_{F, s=\frac{1}{2}}$.\par
Next, we consider the tensor product 
$(d,d)_{B,s=0}\times (d-q,d,q)_{F,s=\frac{1}{2}}$ and look whether, in its irrep decomposition, a leading term of the
form $(d,d)_{s=\frac{3}{2}}$ appears, namely
whether we get
\begin{eqnarray}
(d,d)_{B,s=0}\times (d-q,d,q)_{F,s=\frac{1}{2}}&\equiv&  (d,d)_{s=\frac{3}{2}}+ \ldots
\end{eqnarray}
In this case, the auxiliary fields entering the $(d,d)_{s=\frac{3}{2}}$ irrep on the r.h.s. can be associated with the
(invariant)
kinetic density.\par 
Let us verify how this construction works by computing explicitly the invariant kinetic term for the chiral spin $s=0$ $N=2$ length-$2$ irrep
$(x_1,x_2;\psi_1,\psi_2)$ (in this case $p=q=0$). The needed formulae can be directly read from the $ic$) tensor products 
of (bosonic) multiplets
of Appendix {\bf C} (we use the trick of introducing an $\epsilon$ Clifford parameter to change the statistics of the
$(d,d)_{F, s=\frac{3}{2}}$ fermionic multiplet; this Clifford parameter will be reabsorbed at the end of the computation).
An $N=2$, $(2,2)_{s=\frac{3}{2}}\equiv (\tau_1,\tau_2;w_1,w_2)$ irrep appears on the r.h.s. 
Its component fields are given by
\begin{eqnarray}\label{N2kin}
 \tau_1&=& {\dot x}_1\psi_1-{\dot x}_2\psi_2\nonumber\\
 \tau_2&=& {\dot x}_1\psi_2 +{\dot x}_2\psi_1\nonumber\\
 w_1  &=& {\dot\psi}_1\psi_1+{\dot x}_1{\dot x}_1+{\dot \psi}_2\psi_2+{\dot x}_2{\dot x}_2\nonumber\\
 w_2   &=& {\dot \psi_2}\psi_1+\psi_2{\dot\psi}_1
\end{eqnarray}
Please notice that $w_2$ is a total derivative and therefore does not produce any $N=2$ invariant. On the other
hand, $w_1$ is the required kinetic density, whose time integration provides an $N=2$ invariant action.\par
It should be remarked that the arising of the $(d,d)_{s=\frac{3}{2}}$ term in the r.h.s. of the irrep
decomposition of the $(d,d)_{B,s=0}\times (d-q,d,q)_{F,s=\frac{1}{2}}$ tensor product is not guaranteed and has to
be checked case by case. In particular there is no $N=3$ invariant kinetic term associated with the $N=3$ 
length-$4$ irrep $(1,3,3,1)$ that we discussed at length in the previous two sections. 
\par
So far we have focused ourselves on {\em constant} kinetic terms. 
$\sigma$-model kinetic terms are of more general type (see e.g. \cite{sigma}),
allowing field-dependent metric tensors. An $N$-extended supersymmetric $\sigma$-model invariant action has
the form \par
$S= \int dt \left(g^{ij} (x_k) {\dot x}_i{\dot x}_j + \ldots\right)$, where the $x_i$'s are spin $0$ component fields and the
dots denote the contribution from component fields of spin $s\geq \frac{1}{2}$.
Invariant sigma-models actions can be easily accommodated in our framework. For illustrative purposes we explicitly
discuss the $N=2$ case realized in terms of the previously introduced
spin $s=0$ length-$2$ chiral irrep $(x_1,x_2;\psi_1,\psi_2)$.\par
By tensoring $k$ times the original $(2,2)_{s=0}$ multiplet we get, from the $ic$) case of Appendix {\bf C},
a $k$-linear ${(2,2)_{s=0}}^{(k)}$ multiplet with component fields 
$({x_1}^{(k)}, {x_2}^{(k)};{\psi_1}^{(k)}, {\psi_2}^{(k)})$ 
given by
\begin{eqnarray}\label{N2klin}
{x_1}^{(k)} &=& x_1 {x_1}^{(k-1)} -x_2{x_2}^{(k-1)}\nonumber\\
{x_2}^{(k)}&=&x_1{x_2}^{(k-1)}+x_2{x_1}^{(k-1)}\nonumber\\
{\psi_1}^{(k)}&=&k\psi_1{x_1}^{(k-1)} +k\psi_2{x_2}^{(k-1)}\nonumber\\
{\psi_2}^{(k)}&=& k \psi_2{x_1}^{(k-1)}-k\psi_1{x_2}^{(k-1)}
\end{eqnarray}
A $k$-linear (in the original coordinates) $\sigma$-model type of term is recovered from
the ${(2,2)_{s=\frac{3}{2}}}^{(k)}$ irrep entering the
\begin{eqnarray}
{(2,2)_{s=\frac{3}{2}}}\times{(2,2)_{s=0}}^{(k)}&=&
{(2,2)_{s=\frac{3}{2}}}^{(k)}+\ldots
\end{eqnarray}
decomposition.\par
Up to bilinear products of the original coordinates we get, e.g., the 
following $\sigma$-model type of $N=2$ invariant kinetic term
\begin{eqnarray}\label{sigmaN2}
K&=& 
({{\dot x}_1}^2+{{\dot x}_2}^2-\psi_1{{\dot \psi}_1}-\psi_2{{\dot \psi}_2})\left( C_1(1
+\alpha_1 x_1+\alpha_2({x_1}^2-{x_2}^2))-C_2(\alpha_1x_2+\alpha_2x_1x_2)\right)+\nonumber\\
&&
\psi_1\psi_2\left(C_1(3\alpha_1{\dot x}_2+6\alpha_2({{\dot x}_1}x_2+x_1{{\dot x}_2}))+C_2
(3\alpha_1{\dot x}_1-6\alpha_2({{\dot x}_1}x_1-x_2{{\dot x}_2}))\right)+\ldots\nonumber
\end{eqnarray}
\begin{eqnarray}
&&
\end{eqnarray}
where $\alpha_1$, $\alpha_2$ are arbitrary constants associated to the $k$-linear terms $k=1,2$ respectively,
while $C_1$, $C_2$ are constants related to the two auxiliary fields entering the ${(2,2)_{s=\frac{3}{2}}}^{(k)}$ irrep.

\setcounter{equation}{0}
\section{A class of off-shell invariant actions for general values of $N$}

In this Section we produce a class of off-shell invariant actions for general values $N$ of the $N$-extended
supersymmetries, constructed in terms of the enveloping representations discussed in Section {\bf 5}. To be specific, we discuss at first the invariants associated to the $N$-extended supersymmetry algebra irreps of type $(1;\ldots )$, i.e.  possessing a single, leading,  spin-$0$ bosonic field $x$. The generalization to the other cases will be discussed at the end of this Section.\par
Let us denote with
\begin{eqnarray} \label{symboldef}
x_{i_1\ldots i_k} &\equiv &Q_{i_1} (\ldots  (Q_{i_k} (x)). ) 
\end{eqnarray}
the action of $k$ supersymmetry generators $Q_i$ ($i=1,\ldots, N$) on $x$ (in order to avoid carrying a bunch
of indices, in the following we use the compact notation $x_{i_1\ldots i_k} \equiv x_{[k]}$).\par
By construction, the integral
\begin{eqnarray}\label{adjointf}
{\cal I} &=& \int dt \left(Q_1\cdot \ldots \cdot Q_N f(x)\right)
\end{eqnarray}
applied on a general function $f(x)$, is manifestly $N$-supersymmetric invariant. \par
The integrand $Q_1\ldots Q_N f(x)$ is a lagrangian density of dimension $d=\frac{N}{2}$. It can be proven, in the cases considered below, that it does not coincide with a total derivative\footnote{By general considerations, if for some given $N$, $f(x)$ and choice of the associated irrep, $Q_1\ldots Q_N f(x)$ would coincide with a total derivative, the existence of a non-trivial off-shell invariant of dimension $d<\frac{N}{2}$ would follow as a consequence.}.\par
A bosonic invariant, which would constitute a legitimate off-shell invariant action ${\cal S}$, is recovered for even values of $N$. The supersymmetry generators $Q_i$'s act as graded Leibniz derivatives.
We can represent
${\cal S}$  as
\begin{eqnarray}
{\cal S} &=& \int dt \Omega{\cal B},
\end{eqnarray}
where the action of $\Omega$ on the $x_{[k_1]}^{l_1}\ldots x_{[k]}^{l_k}$ symbols is expressed through
\begin{eqnarray}\label{compactsymbol}
\Omega\left[ x_{[k_1]}^{l_1}\ldots x_{[k_M]}^{l_M}\right] &=& \sum_{perm.} (-1)^{\epsilon_P}\prod_{j=1}^M \left(\frac{1}{k_j!}\right)^{l_j}\frac{1}{l_j!} x_{[k_j]}^{l_j}
\end{eqnarray}
(the sum is over all the permutations and $(-1)^{\epsilon_P}$ is a sign associated to their parity)
and ${\cal B}$ is given, for $N=2,4,6,8,10$, by 
\begin{eqnarray}\label{gradedleibniz}
\begin{array}{|l|l|}\hline
& \quad {\cal B}:\\ \hline
N=2 & f^{(2)} x_{[1]}^2 + f^{(1)}x_{[2]}\\ \hline
N=4 & f^{(4)} x_{[1]}^4 + f^{(3)}x_{[2]}x_{[1]}^2 + 
f^{(2)}[x_{[3]}x_{[1]}+x_{[2]}^2]+f^{(1)}x_{[4]}\\ \hline
N=6 & f^{(6)} x_{[1]}^6 + f^{(5)}x_{[2]}x_{[1]}^4 + 
f^{(4)}[x_{[3]}x_{[1]}^3+x_{[2]}^2x_{[1]}^2]+
f^{(3)}[x_{[4]}x_{[1]}^2+x_{[3]}x_{[2]}x_{[1]}+x_{[2]}^3]+\\ 
&
f^{(2)}[x_{[5]}x_{[1]}+x_{[4]}x_{[2]}+x_{[3]}^2]+
f^{(1)}x_{[6]}\\
\hline
N=8 & f^{(8)} x_{[1]}^8 + f^{(7)}x_{[2]}x_{[1]}^6 + 
f^{(6)}[x_{[3]}x_{[1]}^5+x_{[2]}^2x_{[1]}^4]+
f^{(5)}[x_{[4]}x_{[1]}^4+x_{[3]}x_{[2]}x_{[1]}^3+x_{[2]}^3x_{[1]}^2]+
\\
&
f^{(4)}[x_{[5]}x_{[1]}^3+x_{[4]}x_{[2]}x_{[1]}^2+x_{[3]}^2
x_{[1]}^2+x_{[3]}x_{[2]}^2x_{[1]}+x_{[2]}^4]+\\
&
f^{(3)}[x_{[6]}x_{[1]}^2+x_{[5]}x_{[2]}x_{[1]}+x_{[4]}
x_{[3]}x_{[1]}+x_{[4]}x_{[2]}^2+x_{[3]}^2x_{[2]}]+
\\
&f^{(2)}[x_{[7]}x_{[1]}+x_{[6]}x_{[2]}+x_{[5]}x_{[3]}+ x_{[4]}^2]+
f^{(1)}x_{[8]}\\ \hline
N=10 & f^{(10)} x_{[1]}^{10} + f^{(9)}x_{[2]}x_{[1]}^8 + 
f^{(8)}[x_{[3]}x_{[1]}^7+x_{[2]}^2x_{[1]}^6]+
f^{(7)}[x_{[4]}x_{[1]}^6+x_{[3]}x_{[2]}x_{[1]}^5+x_{[2]}^3x_{[1]}^4]+
\\
&
f^{(6)}[x_{[5]}x_{[1]}^5+x_{[4]}x_{[2]}x_{[1]}^4+x_{[3]}^2
x_{[1]}^4+x_{[3]}x_{[2]}^2x_{[1]}^3+x_{[2]}^4x_{[1]}^2]+\\
&
f^{(5)}[x_{[6]}x_{[1]}^4+x_{[5]}x_{[2]}x_{[1]}^3+x_{[4]}
x_{[3]}x_{[1]}^3+x_{[4]}x_{[2]}^2x_{[1]}^2+x_{[3]}^2x_{[2]}x_{[1]}^2
+x_{[3]}^2x_{[2]}^3x_{[1]}+x_{[2]}^5
]+
\\
&f^{(4)}[x_{[7]}x_{[1]}^3+x_{[6]}x_{[2]}x_{[1]}^2+x_{[5]}x_{[3]}x_{[1]}^2+
x_{[5]}x_{[2]}^2x_{[1]}+x_{[4]}^2x_{[1]}^2+\\
&\quad\quad x_{[4]}x_{[3]}x_{[2]}x_{[1]}+
x_{[4]}x_{[2]}^3+x_{[3]}^3x_{[1]}+
x_{[3]}^2x_{[2]}^2]+\\ 
&f^{(3)}[x_{[8]}x_{[1]}^2+x_{[7]}x_{[2]}x_{[1]}+x_{[6]}
x_{[3]}x_{[1]}+
x_{[6]}x_{[2]}^2\\
&\quad\quad +
x_{[5]}x_{[4]}x_{[1]}+x_{[5]}x_{[3]}x_{[2]}+
x_{[2]}^4x_{[2]}+x_{[4]}x_{[3]}^2]+\\
&
f^{(2)}[x_{[9]}x_{[1]}^2+x_{[8]}x_{[2]}+x_{[7]}x_{[3]}+
x_{[6]}x_{[4]}+x_{[5]}^2]+
f^{(1)}x_{[10]}\\ \hline
\end{array}\nonumber\\
&&
\end{eqnarray}
($f^{(k)}$ denotes the application of $k$ derivatives, $f^{(k)}\equiv \partial_x^kf$).
\par
For $N=2$, the $d=1$ associated invariant corresponds to the most general potential term for the (real) $(1,2,1)$ irreducible
multiplet (the choice $f(x)=\frac{1}{2}x^2$ produces the potential of the $N=2$ harmonic oscillator). \par
For $N=4$,
the $d=2$ associated invariant corresponds to the most general kinetic term for the $(1,4,3)$ irrep, expressed in terms
of an arbitrary function $f(x)$ (the ``constant" kinetic term given by (\ref{kininv}) is recovered for $f(x)=\frac{1}{2}x^2$).
\par
We get in this case, up to integration by parts,
\begin{eqnarray}\label{N4geninvact}
{\cal L} &=& f^{(II)} \left[ {\dot x}^2
-\psi_1{\dot \psi_1}-\psi_2{\dot \psi_2}-\psi_3{\dot \psi_3}-\psi_4{\dot \psi_4}+
{g_1}^2+{g_2}^2+{g_3}^2\right]+ \nonumber\\
&&
f^{(III)} \left[g_1(\psi_1\psi_2+\psi_3\psi_4)
+ g_2(\psi_1\psi_3-\psi_2\psi_4)+
g_3(\psi_1\psi_4+\psi_2\psi_3)\right]+\nonumber\\
&&
 f^{(IV)}\psi_1\psi_2\psi_3\psi_4 \nonumber\\
\end{eqnarray}
The same construction can be iterated for $N=6,8,10,\ldots$ . The complete list of spin $s=0$ irreducible multiplets
of type $(1;\ldots )$, taken from the Appendices ${\bf A}$ and ${\bf B}$, is given by
\begin{eqnarray}\label{oneplusextramult}
 \begin{array}{|l|l|l|}\hline
N=2 & 1& (1,2,1) \\ \hline
N=4 & 1&(1,4,3)\\ \hline
N=6 & 3&(1,8,7), (1,7,7,1), (1,6,7,2)\\ \hline
N=8 & 1&(1,8,7) \\ \hline
N=10 &1+22+49=72&length-3: (1,32,31)\\ 
& & length-4: (1,32-k,31, k) \quad for \quad k=1,\ldots ,22\\ 
&&length-5: (1,10+n, 31-m, 22-n, m)\quad for\\ 
&& \quad \quad\quad\quad\quad m~positive, \quad n ~non-negative \quad and \quad
 n+2m\leq 14\\ \hline
\end{array}\nonumber\\
\end{eqnarray}
For $N=10$ we obtain from the above construction $72$ inequivalent off-shell invariant actions, 
associated with each given irrep. All these invariants depend on a general function $f(x)$.\par
The increment of computational work required for larger values of $N$ can be measured by the number of terms,
ordered by increasing derivatives of $f$, entering the
second column of (\ref{gradedleibniz}). Applying some combinatorics we get 
 \begin{eqnarray}\label{numberofterms}
\begin{array}{|l|ll|}\hline
N=2 & 1+1 &=2 \\ \hline
N=4 & 1+7+6+1&=15\\ \hline
N=6 & 1+31+90+65+15+1&=203\\
\hline
N=8 & 1+127+ 966+3409+1330+266+28+1 &=6128 \\ \hline
N=10 & 1+511+11745+34105+42525+22827+5880+750+45+1&=118390\\ \hline
\end{array}\nonumber\\
\end{eqnarray} 
We give an example of $N=6$ invariant action for one of the three available irreps. Let us take 
the length-$4$ $(1,7,7,1)\equiv (x; \psi_i; g_i;\omega)$, for $i=1,\ldots, 7$, irrep (see Appendix {\bf A}).\par
We obtain
\begin{eqnarray}\label{N6action}
{\cal L} &=& f^{(I)}{\ddot g}_7 + \nonumber\\
&& f^{(II)}[
{\ddot \psi}_1\psi_6 +2
{\dot \psi}_6{\dot \psi}_1 +\psi_1{\ddot \psi}_6+
{\ddot \psi}_2\psi_5 +2
{\dot \psi}_5{\dot \psi}_2 +\psi_2{\ddot \psi}_5+
{\ddot \psi}_4\psi_3 +2
{\dot \psi}_3{\dot \psi}_4 +\psi_4{\ddot \psi}_3
+\nonumber \\&&4\omega {\dot \psi}_7+
2(g_1{\dot g}_6- g_6{\dot g}_1) + 2(g_2{\dot g}_5- g_5{\dot g}_2)+
2(g_4{\dot g}_3- g_3{\dot g}_4)+ 3 g_7 {\ddot x}
] + \left( \ldots \right)\nonumber\\
\end{eqnarray}
The dots in the r.h.s. denote the terms presenting at least a third-order derivative of $f$. The terms explicitly present
in (\ref{N6action}) are sufficient to compute the invariant action associated with $f(x)=\frac{1}{2}x^2$.
We get, up to integration by parts, a lagrangian ${\cal L}$
\begin{eqnarray}\label{N6action2}
\frac{1}{4}{\cal L} &=& g_7{\ddot x} +
g_1{\dot g}_6 +
g_2{\dot g}_5 -
g_3{\dot g}_4 +
\omega{\dot \psi_7} + \psi_1{\ddot \psi}_6+
\psi_2{\ddot \psi}_5-\psi_3{\ddot \psi}_4
\end{eqnarray}
The results for higher values of $N$ ($N=8,10,\ldots$) are too cumbersome to be explicitly written in terms
of the component fields entering the multiplets.\par
At this point some remarks are in order.
The ``embedding" of (the general function of component fields of) an irrep into the enveloping representation automatically produces non-trivial off-shell invariants. The construction presented above can be straightforwardly generalized to
any type of irreps and for several classes of functions. An irrep of the $N$-extended supersymmetry with $p$ leading spin-$0$ fields $x_1, \ldots, x_p$,
produces an invariant by applying $Q_1\ldots Q_N$ to the $0$-dimensional function of $p$ variables $f(x_1,\ldots, x_p)$.
Invariants are also obtained by applying $Q_1\ldots Q_N$ to functions of dimension $\frac{1}{2}$, $1$, etc.
As an example, the $d=2$ kinetic term ${\cal L}_2$ of the real $N=2$ $(1,2,1)\equiv (x;\psi_1,\psi_2;g)$ multiplet is given
by applying $Q_1,Q_2$ to a dimension $d=1$ function, 
\begin{eqnarray}
{\cal L}_2&=& Q_1Q_2 \left( a(x)\psi_1\psi_2+b(x)g\right).
\end{eqnarray}
Explicitly, we get
\begin{eqnarray}\label{N2kinetic}
{\cal L}_2 &=& \mu(x) \left[ {\dot x}^2 +g^2-\psi_1{\dot \psi}_1-\psi_2{\dot\psi}_2\right]-\mu'\psi_1\psi_2g
\end{eqnarray}
with $\mu(x) = a(x)+b'(x)$.\par  
Similarly, the ${\widetilde{\cal L}}_1$ $d=1$ and ${\widetilde{\cal L}}_2$ $d=2$ invariant densities of the
$N=2$ (chiral) $(2,2)\equiv (x_1,x_2;\psi_1,\psi_2)$ multiplet are respectively obtained  through
\begin{eqnarray}\label{N2chiral}\begin{array}{lllll}
{\widetilde{\cal L}}_1 &=& Q_1Q_2f(x_1,x_2)&=& -2\partial_1\partial_2 f \cdot \psi_1\psi_2\\
{\widetilde{\cal L}}_2 &=& Q_1Q_2 \gamma(x_1,x_2) \cdot\psi_1\psi_2 &=  &\gamma(x_1,x_2)\left[{\dot x}_1^2+{\dot x}_2^2 -\psi_1{\dot\psi}_1-\psi_2{\dot\psi}_2
\right] +\\
&&&&\psi_1\psi_2({\dot x}_1\partial_2\gamma-{\dot x}_2\partial_1\gamma)
\end{array}
\end{eqnarray}
This generalization can be used to construct bosonic off-shell invariants for odd-$N$ supersymmetries. 
Bosonic $N=9$ invariants for the $N=9$ irreps can, e.g., be obtained by applying $Q_1\ldots Q_9$ to the $d=\frac{1}{2}$ function $\sum_j f_j\psi_j$, where the sum is extended over all the spin $\frac{1}{2}$ fields $\psi_j$ entering the 
given irrep
and $f_j$ are arbitrary functions of the spin-$0$ fields.  \par
For large values of $N$ the off-shell actions produced with our construction have large dimensions
and involve higher-order time derivatives. It seems plausible that at least some of these invariant actions could arise from  the dimensional reduction of terms entering supersymmetric field theories
with higher derivatives, such as Born-Infeld actions.\par
Let us conclude this section by pointing out that the class of ``enveloping" off-shell invariants here discussed
does not exhaust the whole class of off-shell invariants. In the previous Section, e.g., we proved, see (\ref{s1twoinv}),
that $N=4$ bilinear and multi-linear invariants of the $(2,4,2)$ multiplet are encountered due to the fact that the tensor product of two $(2,4,2)$ irreps produces (formula (\ref{242tensor}))  another $(2,4,2)$ multiplet in the r.h.s .
These invariants have spin $s=1$, while the ``enveloping construction" above produces, for $N=4$,
invariants of spin $s\geq 2$. \par
These invariants, of ``lesser spin", are the result of a Clebsch-Gordan type of decomposition of irreps. For larger values of $N$, a still open question concerns the decomposition of the tensor product of irreps into the fundamental irreps.
This is a technical problem, still unsolved due to the large computational work that it requires.\par
The method here developed allows us to get at least partial answers. We explicitly discuss here the $N=8$ off-shell
invariant action of dimension $s=2$ for the $(1,8,7)$ multiplet. 

\subsection{The $N=8$ off-shell invariant action of the $(1,8,7)$ multiplet}

It is convenient at first to rewrite the most general $N=4$ off-shell invariant action (\ref{N4geninvact}) of the $(1,4,3)$ multiplet 
by making explicit that it can be covariantly written in terms of the tensor $\delta_{ij}$ and 
the totally antisymmetric tensor $\epsilon_{ijk}$ ($i,j,k=1,2,3$, $\epsilon_{123}=1$)
acting on the three imaginary quaternions. Indeed, the three auxiliary fields $g_i$'s and three of the four fermionic
fields $\psi$'s can be associated to the three imaginary quaternions. The $N=4$ supersymmetry transformations can
therefore be written
as
\begin{eqnarray}
Q_i (x; \psi, \psi_j, g_j) &=& (-\psi_i; g_i, -\delta_{ij}{\dot x} +\epsilon_{ijk} g_k; \delta_{ij}{\dot\psi}-\epsilon_{ijk}
{\dot\psi}_k),\nonumber\\
Q_4 ( x; \psi, \psi_j; g_j) &=& (\psi; {\dot x}, g_j;{\dot \psi}_j)
\end{eqnarray}
(the Einstein convention over repeated indices is understood). The off-shell action (\ref{N4geninvact})
now reads
\begin{eqnarray}\label{covariantN4action}
\relax{\cal S} &=& \int dt \{\alpha(x)[{\dot x}^2 -\psi{\dot\psi} -\psi_i{\dot\psi}_i +{g_i}^2]+\nonumber\\
&&
\alpha'(x)[ \psi g_i\psi_i -\frac{1}{2}\epsilon_{ijk}g_i\psi_j\psi_k] 
-\frac{\alpha''(x)}{6}[\epsilon_{ijk}\psi\psi_i\psi_j\psi_k]\}
\end{eqnarray}
where $\alpha(x)$ is an arbitrary function.\par
The strategy to construct the corresponding, most general $N=8$ off-shell invariant action of the $(1,8,7)$ multiplet
is now clear. From the discussion in Section {\bf 3} we know that the $N=8$ supersymmetry can be produced from
the lifting of the $Cl(0,7)$ Clifford algebra to $Cl(9,0)$. On the other hand it is well-known \cite{crt2}, that the
seven $8\times 8$ antisymmetric gamma matrices of $Cl(0,7)$ can be recovered by the left-action of the imaginary
octonions on the octonionic space. As a result, the entries of the seven antisymmetric gamma-matrices of
$Cl(0,7)$ can be expressed in terms of the totally antisymmetric octonionic structure constants $C_{ijk}$'s.
The non-vanishing $C_{ijk}$'s are given by
\begin{eqnarray}
&C_{123}=C_{147}=C_{165}=C_{246}=C_{257}=C_{354}=C_{367}=1&
\end{eqnarray}
The $N=8$ supersymmetry transformations of the $(1,8,7)$ multiplet are expressed as
\begin{eqnarray}
Q_i (x; \psi, \psi_j, g_j) &=& (-\psi_i; g_i, -\delta_{ij}{\dot x} +C_{ijk} g_k; \delta_{ij}{\dot\psi}-C_{ijk}
{\dot\psi}_k),\nonumber\\
Q_8 ( x; \psi, \psi_j; g_j) &=& (\psi; {\dot x}, g_j;{\dot \psi}_j)
\end{eqnarray}
for $i,j,k=1,\ldots, 7$.
We can therefore look for the most general $N=8$ invariant action, covariantly written in terms of the octonionic
structure constants and which reduces to (\ref{covariantN4action}) when restricted to the quaternionic
(i.e. $N=4$) subspace. With respect to (\ref{covariantN4action}), an extra-term could be in principle present.
It is given by $\int dt \beta(x)C_{ijkl}\psi_i\psi_j\psi_k\psi_l$ and is constructed in terms of the octonionic tensor of
rank $4$ 
\begin{eqnarray}
C_{ijkl}&=&\frac{1}{6}\epsilon_{ijklmnp}C_{mnp}
\end{eqnarray}
(where $\epsilon_{ijklmnp}$ is the seven-indices totally antisymmetric tensor).
Please notice that the rank-$4$ tensor is obviously vanishing when restricting to the quaternionic subspace.
One immediately verifies that the term $\int dt \beta(x)C_{ijkl}\psi_i\psi_j\psi_k\psi_l$ breaks the $N=8$
supersymmetries and cannot enter an invariant action. For what concerns the other terms, starting from the general
action (with $i,j,k=1,\ldots, 7$)
\begin{eqnarray}\label{covariantN8action}
\relax{\cal S} &=& \int dt \{\alpha(x)[{\dot x}^2 -\psi{\dot\psi} -\psi_i{\dot\psi}_i +{g_i}^2]+\nonumber\\
&&
\alpha'(x)[ \psi g_i\psi_i -\frac{1}{2}C_{ijk}g_i\psi_j\psi_k] 
-\frac{\alpha''(x)}{6}[C_{ijk}\psi\psi_i\psi_j\psi_k]\}
\end{eqnarray}
we can prove that the invariance under the $Q_i$ generator ($=1,\ldots 7$) is broken by terms which, after integration by parts,
contain at least a second derivative $\alpha''$. We obtain, e.g., a non-vanishing term of the
type $\int dt \alpha''\frac{\psi}{2}C_{ijkl}g_j\psi_k\psi_l$. In order to guarantee the full $N=8$ invariance
(the invariance under $Q_8$ is automatically guaranteed)
we have therefore to set $\alpha''(x)=0$, leaving $\alpha$ a linear function in $x$.
The most general $N=8$ off-shell invariant action of the $(1,8,7)$ multiplet is given by
\begin{eqnarray}\label{N8invact}
{\cal S} &=& \int dt \{(ax +b) [{\dot x}^2 -\psi{\dot\psi} -\psi_i{\dot\psi}_i +{g_i}^2]+ a
[ \psi g_i\psi_i -\frac{1}{2}C_{ijk}g_i\psi_j\psi_k] \}
\end{eqnarray}
Some remarks are in order. The only sign of the octonions is through their structure constants entering as
parameters in the (\ref{N8invact}) $N=8$ off-shell invariant action. This means that (\ref{N8invact}) is an ordinary
action, in terms of ordinary bosonic and fermionic fields closing an ordinary $N=8$ supersymmetry algebra.
The procedure here outlined, covariantization w.r.t. the octonionic structure constants, can be repeated for other
multiplets and for other values of the $N$-extended supersymmetries. The results will be reported elsewhere.
\par
Summarizing, we have produced partial, but new results in the classification of off-shell invariant actions.
On the other hand, due to the results presented here for the first time, we finally possess the complete classification of irreps for $N\geq 9$ (only partial results were previously known in the literature).
\par
For not so large values of $N$ (possibly $N=9,10,\ldots$), a ``brute force" method to construct all invariants, that we are leaving
for future investigations, could perhaps prove useful. For (arbitrarily) large values of $N$, new concepts are required. In the next Section we discuss a class of useful information, encoded in the ``fusion algebra" of the supersymmetric vacua irreps.

\setcounter{equation}{0}
\section{The fusion algebra of the supersymmetric vacua}

The supersymmetry transformations of the component fields in a multiplet involve time derivatives.
When tensoring, e.g., two irreps, the resulting representation can be decomposed into its irreducible constituents.
This, in general, will produce infinite towers of irreps of increasing spin, in terms of bilinear products of the original
component fields and their (higher-order) derivatives. However, a drastic simplification arises when we 
consistently set all time derivatives equal to zero. Since the hamiltonian operator acts, up to a factor, as a time derivative,
this is tantamount to analyze the decomposition of the tensored products of irreps at the zero-energy level,
i.e. for the unbroken supersymmetric vacua.\par
As an example, when tensoring two real $N=2$ spin $s=0$ irreps (we remember that each one is given by two spin $s=0$ and two spin $s=\frac{1}{2}$ component fields),
we obtain $4$ spin $s=0$, $8$ spin $s=\frac{1}{2}$ and $4$ spin $s=1$ bilinear fields entering a reducible, vacuum ($0$-energy)
supersymmetric multiplet.\par
By tensoring supersymmetric vacua irreps we always produce a finite number of fields. As a consequence, we are allowed to introduce a notion of fusion algebra for the supersymmetric vacua, which can be mimicked after the notion of fusion algebra for the rational conformal field theories (RCFT), see e.g. \cite{gab}.\par
The fusion algebra can be defined as follows. Let us denote with $[i]$, $[j]$, $[k]$, $\ldots$ the inequivalent irreps
of an $N$-extended supersymmetry, 
where $i,j,k$ takes the values $1,2,\ldots , {\overline\kappa}$ or 
$1,2,\ldots, {\kappa}=2{\overline\kappa}$ (see table (\ref{inequivalent}) and the discussion thereafter), according to
whether we discriminate or not irreps according to their bosonic/fermionic character.\par
The tensoring of two zero-energy vacuum-state irreps can be symbolically written as
\
\begin{eqnarray}\label{fusion}
\relax [i]\times [j] &=& {N_{ij}}^k [k]
\end{eqnarray}
where
${N_{ij}}^k$ are non-negative integers specifying the decomposition of the tensored-products irreps into its irreducible constituents. The ${N_{ij}}^k$ integers satisfy a fusion algebra with the following properties\par
$1$) Constraint on the total number of component fields,
\begin{eqnarray}\label{ineqconstr}
\forall ~ i,j\quad \sum_k {N_{ij}}^k &=&2d
\end{eqnarray} 
where $d$ (see (\ref{irrepdim})) is the number of bosonic (fermionic) fields
in the given irreps. Please notice that, for a fixed $N$-extended supersymmetry, 
$d$ is the same for any irrep.\par
$2$) The symmetry property 
\begin{eqnarray}\label{fusionsymm}
{N_{ij}}^k&=&{N_{ji}}^k
\end{eqnarray}

$3$) The associativity condition. This property can be expressed on the ${N_{ij}}^k$ integers as follows. Let us set ${{(N_i)}_j}^k =_{def} {N_{ij}}^k$, then
the relation 
\begin{eqnarray}\label{assoc}
\relax [i]\times ([j]\times[k]) &=&([i]\times[j])\times[k]
\end{eqnarray}
implies that the
r.h.s. $\relax {N_{ij}}^r[r]\times[k]= {N_{ij}}^r{N_{rk}}^t[t]={(N_iN_k)_j}^t[t]$ 
coincides with the l.h.s.
$\relax [i]\times{N_{jk}}^r[r]={N_{jk}}^r{N_{ir}}^t[t]= {(N_kN_i)_j}^t[t]$. Namely, the associativity
condition (\ref{assoc}) implies the commutativity of the fusion matrices
\begin{eqnarray}\label{commut}
[N_i,N_k]&=& 0
\end{eqnarray}
The notion of fusion algebra of the supersymmetric vacua can be usefully applied even when tensoring fundamental irreps
that do not satisfy the vacuum condition. For instance, the decomposition into irreps of a leading tensored multiplet (see appendix ${\bf C}$ for a discussion) can be directly read from the associated vacuum fusion algebra.\par
It is worth mentioning that the above properties of the ${N_{ij}}^k$ fusion matrices can, in some cases, help determining
the decomposition into irreps without explicitly computing them. The simplest such kind of application is discussed in appendix ${\bf D}$, where the fusion algebras (with and without discrimination of the bosonic/fermionic character) of the $N=2$ supersymmetric vacua are explicitly reported.   

\newpage

\setcounter{equation}{0}

\section{Conclusions}

In this paper we combined the results of \cite{pt}, \cite{crt2} and \cite{kt} in order to produce the classification of the
linear, finite, irreducible representations of the $N$-extended one-dimensional supersymmetry algebras. The \cite{crt2} algorithmic construction of Clifford algebras was used to compute, for each given value $N$, the irreducible representations.
The complete classification has been here explicitly presented up to $N\leq 10$, while the length-$4$ irreps have been
reported for the oxidized $N=11^{\ast}, 12$ extended supersymmetries as well. We proved that the \cite{kt} results on oxidized Clifford
algebras imply, as a corollary, that the $N=3,5\quad mod\quad 8$ extended supersymmetries admit two classes of
irreps, real and quaternionic, while the remaining values of $N$ admit a unique class of irreps.\par
We further produced tensorings of irreps and showed how to use them in order to construct manifest multi-linear invariants of the
$N$-extended supersymmetries with no need of introducing an associated superspace formalism. We pointed out that the invariants
can be realized either in terms of {\em unconstrained} fields entering an irreducible multiplet or even, in specific cases,
consistently {\em multilinearly
constrained} fields. \par 
A whole class of off-shell invariants for arbitrarily large values $N$ of the extended supersymmetry and for any
given associated irreducible representation have been constructed in Section {\bf 7}. 
In the same section we also produced the most general $N=8$ invariant off-shell action
of the $(1,8,7)$ multiplet, via the ``octonionic-covariantization technique". The complete classification of off-shell invariants,
as reminded in Section {\bf 7}, requires solving, for large values of $N$, the decomposition into irreps of the irreps tensor
products. This problem, still intractable for $N\geq 9$, can start being addressed with the information contained 
in the so called {\em fusion algebra} of the supersymmetric vacua. In the last Section we
introduced the notion of the {\em fusion algebra} of the supersymmetric vacua. We explicitly presented its simplest
non-trivial (for $N=2$) example in Appendix {\bf D}.\par
In the Introduction we have already discussed several possible applications of the present results, for instance the classification of one-dimensional sigma-models, see e.g. \cite{sigma}, admitting $N$-extended supersymmetries. Another class of models which could
be profitably investigated within this framework concerns the integrable systems in $1+1$ dimensions with extended number of supersymmetries,
see \cite{gr2} and \cite{crt1}. \par
It is worth mentioning another line of development which is outside the scope of the present paper and deserves further investigation. It concerns the systematic 
construction of the non-linear realizations of the extended supersymmetries. Some remarks about the relation
between linear representations and non-linear realizations of the $N$-extended supersymmetry algebra can be found, e.g., in
\cite{abc}. The most recent papers on non-linear Supersymmetric Mechanics are \cite{nlin} (see also the references therein).\par
Let us finally conclude this paper by pointing out that it would be quite appealing to implement the algorithms here
discussed in a computer algebra package. This would allow to explicitly produce for {\em arbitrarily}
(in practice, the limit is due to the available computation time) large 
values of $N$ the complete list of inequivalent irreps (the results explicitly reported here have been derived without any
computer help). Such a package, once implemented, would allow to perform mathematical experiments which could lead to
conjecture possible closed formulas satisfied by irreps for arbitrarily large values of $N$.

\renewcommand{\theequation}{A.\arabic{equation}}
\setcounter{equation}{0}
\par{~}\\
{\Large{\bf Appendix A \\{~}\\
Supersymmetry irreps for $N\leq 8$}}\par
{~}\par
We focus here on the irreducible multiplets with length $l\leq 4$ since, applying the method described
in Section {\bf 4}, we proved that the extended supersymmetries
with $N\leq 9$ do not admit irreps with length higher than $4$. We denote $l\leq 4$ irreps 
as $(d-p, d-q, p,q)$, where $d$ is the total number of bosonic (fermionic) fields
entering the multiplet. According to the results of Section {\bf 2}, the length
$2$ multiplets correspond to $p=q=0$, while the length $3$ multiplets are recovered
from $q=0$, $p\neq0$ ($p=1,2,\ldots , d-1$).
\par
As recalled in section {\bf 2}, irrep multiplets are either bosonic or fermionic according to
the statistics of their leading component fields (the fields with lowest spin, whose total number
is given by $d-p$). For our purposes it is convenient to denote as $x_i$, ($i=1,\ldots , d-p$)
such leading component fields. Their spin is conventionally assigned to be
$s$. The $d-q$ fields of spin $s+\frac{1}{2}$ are here denoted as 
$\psi_j$ ($j=1,\ldots , d-q$), while the $p>0$ fields of spin $s+1$ are expressed as
 $g_k$ ($k=1, \ldots,
p$). The $q$ fields of spin $s+\frac{3}{2}$ (for $q>0$) are in the following denoted as $\omega_l$, $l=1,\ldots, q$. 
Therefore $(d-p,d-q,p,q) \equiv (x_i;\psi_j;g_k;\omega_l)$.\par
The fields $x_i, g_k$ are all bosonic (fermionic) and the fields $\psi_j, \omega_l$ are all fermionic (bosonic)
if the associated multiplet is bosonic (fermionic). It should be noticed that the transformation properties
of the fields entering the multiplet do not depend on the statistics (either bosonic or fermionic) of the multiplet.
Therefore, in the following we do not need to specify whether the irreducible multiplets under consideration are bosonic or fermionic. \par
All irreps of the $N$-extended supersymmetry can be systematically computed (for any arbitrary value $N$) with the
algorithmic construction presented in Sections {\bf 2-4}. For completeness, it is convenient to explicitly present in this appendix
all irreps up to $N=8$, furnishing a representative in each irreducible class. For $N=3,5$ two classes of inequivalent irreps, real and quaternionic, have to be presented in accordance with the results of Section {\bf 3}.\par
We get the following list of irreps
\par{~}\par
{\em i}) {\bf The $N=1$ irrep}\par
We have only one irrep, $(1,1)\equiv (x;\psi)$, with transformation property
\begin{eqnarray}
Q_1 (1,1) &=& (\psi; {\dot x}).
\end{eqnarray}
\par
{~}\par
{\em ii}) {\bf The $N=2$ irreps}\par
There are two inequivalent irreps\par
\begin{eqnarray}
 (2,2)&\equiv& (x_1,x_2;\psi_1,\psi_2),\nonumber\\
 (1,2,1)&\equiv & (x; \psi_1,\psi_2; g),
\end{eqnarray}
whose respective supersymmetry transformations are given by
\begin{eqnarray}
Q_1(2,2) &=& (\psi_2, \psi_1; {\dot x}_2,{\dot x}_1)\nonumber\\
Q_2(2,2) &=& (\psi_1, -\psi_2; {\dot x}_1,-{\dot x}_2)
\end{eqnarray}
and
\begin{eqnarray}
Q_1(1,2,1) &=& (\psi_2; g, {\dot x}; {\dot\psi}_1)\nonumber\\
Q_2(1,2,1) &=& (\psi_1; {\dot x}, -g; -{\dot\psi}_2)
\end{eqnarray}

\par
{~}\par
{\em iii}) {\bf The real $N=3^{(\ast)}$ irreps}\par
They are recovered from the $Cl(4,3)$ Clifford algebra (see Section {\bf 3}).
This case admits $5$ inequivalent irreps, labeled as
\begin{eqnarray}
 (4,4) &\equiv& (x_1,x_2, x_3, x_4;\psi_1,\psi_2,\psi_3,\psi_4),\nonumber\\
 (3,4,1) &\equiv & (x_1, x_2, x_3; \psi_1,\psi_2, \psi_3, \psi_4; g),\nonumber\\
 (2,4,2) &\equiv& (x_1,x_2;\psi_1,\psi_2, \psi_3,\psi_4;g_1,g_2),\nonumber\\
 (1,4,3) &\equiv & (x; \psi_1,\psi_2,\psi_3,\psi_4; g_1,g_2,g_3),\nonumber\\
 (1,3,3,1) &\equiv & (x; \psi_1,\psi_2,\psi_3; g_1,g_2,g_3; \omega)
\end{eqnarray}
Their supersymmetry transformations are respectively given by
\begin{eqnarray}
Q_1(4,4) &=& (\psi_4, \psi_3,\psi_2, \psi_1; {\dot x}_4, {\dot x}_3, {\dot x}_2, {\dot x}_1)\nonumber\\
Q_2(4,4) &=& (\psi_3, -\psi_4,\psi_1, -\psi_2; {\dot x}_3, -{\dot x}_4, {\dot x}_1, -{\dot x}_2)\nonumber\\
Q_3(4,4) &=& (\psi_1, \psi_2,-\psi_3, -\psi_4; {\dot x}_1, {\dot x}_2, -{\dot x}_3, -{\dot x}_4)
\end{eqnarray}
\begin{eqnarray}
Q_1(3,4,1) &=& (\psi_4, \psi_3, \psi_2; g, {\dot x}_3, {\dot x}_2, {\dot x}_1;{\dot \psi}_1)\nonumber\\
Q_2(3,4,1) &=&  (\psi_3, -\psi_4, \psi_1; {\dot x}_3, -g, {\dot x}_1,- {\dot x}_2;-{\dot \psi}_2)\nonumber\\
Q_3(3,4,1) &=&  (\psi_1, \psi_2, -\psi_3; {\dot x}_1, {\dot x}_2,-{\dot x}_3, -g;-{\dot \psi}_4)
\end{eqnarray}
\begin{eqnarray}
Q_1(2,4,2) &=& (\psi_4, \psi_3; g_2, g_1, {\dot x}_2, {\dot x}_1; {\dot \psi}_2, {\dot \psi}_1)\nonumber\\
Q_2(2,4,2) &=& (\psi_3, -\psi_4; g_1, -g_2, {\dot x}_1, -{\dot x}_2; {\dot\psi}_1, -{\dot \psi}_2)\nonumber\\
Q_3(2,4,2) &=& (\psi_1, \psi_2; {\dot x}_1, {\dot x}_2, -g_1,-g_2; -{\dot\psi}_3, -{\dot \psi}_4)
\end{eqnarray}
\begin{eqnarray}
Q_1(1,4,3) &=& (\psi_4; g_3, g_2, g_1, {\dot x}; {\dot \psi}_3, {\dot \psi}_2, {\dot \psi}_1)\nonumber\\
Q_2(1,4,3) &=& (\psi_3; g_2, -g_3, {\dot x}, -g_1; -{\dot \psi}_4, {\dot\psi}_1, -{\dot \psi}_2)\nonumber\\
Q_3(1,4,3) &=& (\psi_1; {\dot x}, g_1, -g_2, -g_3; {\dot\psi}_2, -{\dot\psi}_3, -{\dot \psi}_4)
\end{eqnarray}
and, for the length-$4$ irrep,
\begin{eqnarray}
Q_1(1,3,3,1) &=& (\psi_3; g_3, g_1, {\dot x}; {\dot \psi}_2, \omega, {\dot \psi}_1;{\dot g}_2)\nonumber\\
Q_2(1,3,3,1) &=& (\psi_2; g_2,{\dot x}, -g_1; -{\dot \psi}_3, {\dot\psi}_1, -\omega;-{\dot g}_3)\nonumber\\
Q_3(1,3,3,1) &=& (\psi_1;{\dot x}, -g_2, - g_3;\omega, -{\dot\psi}_2, -{\dot\psi}_3; {\dot g}_1)
\end{eqnarray}
\par
{~}\par
{\em iv}) {\bf The quaternionic $N=3^{(\ast\ast)}$ irreps}\par
The length $2$ and $3$ irreps of the quaternionic $N=3$ supersymmetry
can be directly read from the transformations of the $N=4$ irreps
(since $N=4$ is the {\em oxidized} supersymmetry of the quaternionic $N=3$), by restricting the 
supersymmetry transformations
to be given by $Q_1$, $Q_2$, $Q_3$.\par
An extra, length-$4$, irrep
is given by 
\begin{eqnarray}
(1,3,3,1) &\equiv& (x; \psi_1, \psi_2,\psi_3; g_1, g_2, g_3;\omega).
\end{eqnarray}
Its supersymmetry transformations are
\begin{eqnarray}
Q_1(1,3,3,1) &=& (\psi_1; {\dot x}, g_3, -g_2; -\omega, -{\dot \psi}_3, {\dot \psi}_2;-{\dot g}_1)\nonumber\\
Q_2(1,3,3,1) &=& (\psi_3; g_2, -g_1, {\dot x}; -{\dot \psi}_2, {\dot\psi}_1, -\omega;-{\dot g}_3)\nonumber\\
Q_3(1,3,3,1) &=& (\psi_2; -g_3, {\dot x}, g_1; {\dot\psi}_3, -\omega, -{\dot\psi}_1; -{\dot g}_2)
\end{eqnarray}
\par
{~}\par
{\em v}) {\bf The $N=4$ irreps}\par
This case admits $4$ inequivalent irreps, \par
\begin{eqnarray}
 (4,4) &\equiv& (x_1,x_2, x_3, x_4;\psi_1,\psi_2,\psi_3,\psi_4),\nonumber\\
 (3,4,1) &\equiv & (x_1, x_2, x_3; \psi_1,\psi_2, \psi_3, \psi_4; g),\nonumber\\
 (2,4,2) &\equiv& (x_1,x_2;\psi_1,\psi_2, \psi_3,\psi_4;g_1,g_2),\nonumber\\
 (1,4,3) &\equiv & (x; \psi_1,\psi_2,\psi_3,\psi_4; g_1,g_2,g_3),
\end{eqnarray}
whose supersymmetry transformations are respectively given by
\begin{eqnarray}
Q_1(4,4) &=& (\psi_2, -\psi_1, -\psi_4, \psi_3; -{\dot x}_2,{\dot x}_1, {\dot x}_4, -{\dot x}_3)\nonumber\\
Q_2(4,4) &=& (\psi_4, -\psi_3, \psi_2, -\psi_1; -{\dot x}_4,{\dot x}_3, -{\dot x}_2, {\dot x}_1)\nonumber\\
Q_3(4,4) &=& (\psi_3, \psi_4, -\psi_1, -\psi_2; -{\dot x}_3,-{\dot x}_4, {\dot x}_1, {\dot x}_2)\nonumber\\
Q_4 (4,4) &=& (\psi_1, \psi_2, \psi_3, \psi_4; {\dot x}_1,{\dot x}_2, {\dot x}_3, {\dot x}_4)
\end{eqnarray}
\begin{eqnarray}
Q_1(3,4,1) &=& (\psi_2, -\psi_1, -\psi_4;  -{\dot x}_2,{\dot x}_1,g,- {\dot x}_3; {\dot \psi}_3)\nonumber\\
Q_2(3,4,1) &=& (\psi_4, -\psi_3, \psi_2; -g, {\dot x}_3,-{\dot x}_2, {\dot x}_1;- {\dot \psi}_1)\nonumber\\
Q_3(3,4,1) &=& (\psi_3, \psi_4, -\psi_1;  -{\dot x}_3,-g, {\dot x}_1, {\dot x}_2;- {\dot \psi}_2)\nonumber\\
Q_4 (3,4,1) &=& (\psi_1, \psi_2, \psi_3;  {\dot x}_1,{\dot x}_2, {\dot x}_3, g; {\dot \psi}_4)
\end{eqnarray}
\begin{eqnarray}
Q_1(2,4,2) &=& (\psi_2, -\psi_1;  -{\dot x}_2,{\dot x}_1,g_2, -g_1; -{\dot \psi}_4, {\dot \psi}_3)\nonumber\\
Q_2(2,4,2) &=& (\psi_4, -\psi_3; -g_2, g_1, -{\dot x}_2,{\dot x}_1; {\dot \psi}_2, -{\dot \psi}_1)\nonumber\\
Q_3(2,4,2) &=& (\psi_3, \psi_4; -g_1, -g_2, {\dot x}_1,{\dot x}_2;- {\dot \psi}_1, -{\dot \psi}_2)\nonumber\\
Q_4 (2,4,2) &=& (\psi_1, \psi_2;  {\dot x}_1,{\dot x}_2,g_1,g_2; {\dot \psi}_3, {\dot \psi}_4)
\end{eqnarray}
\begin{eqnarray}
Q_1(1,4,3) &=& (\psi_2;-g_1, {\dot x}, g_3, -g_2;  -{\dot \psi}_1,-{\dot \psi}_4, {\dot \psi}_3)\nonumber\\
Q_2(1,4,3) &=& (\psi_4;-g_3, g_2, -g_1, {\dot x};  -{\dot \psi}_3,{\dot \psi}_2, -{\dot \psi}_1)\nonumber\\
Q_3(1,4,3) &=& (\psi_3;-g_2, -g_3, {\dot x}, g_1;  {\dot \psi}_4,-{\dot \psi}_1, -{\dot \psi}_2)\nonumber\\
Q_4 (1,4,3) &=& (\psi_1;{\dot x}, g_1, g_2, g_3;  {\dot \psi}_2,{\dot \psi}_3, {\dot \psi}_4)
\end{eqnarray}

\par
{~}\par
{\em vi}) {\bf The real $N=5^{(\ast)}$ irreps}\par
The length $2$ and $3$ irreps are obtained from the $N=8$ irreps by restricting
the supersymmetry transformations to be given by $Q_i$, for $i=1,\ldots, 5$.
\par
This case admits two, dually related (see Section {\bf 4}), length-$4$ irreps
which cannot be oxidized to $N=6,7$ irreps and four extra irreps, three of them
oxidized to $N=6$, while the last irrep is oxidized to $N=7$. These extra four irreps
are presented in the following. The two length-$4$ irreps with maximal number
of $N=5^{(\ast)}$ supersymmetries are given by 
\begin{eqnarray}
(1,5,7,3) &=& (x; \psi_1,\ldots, \psi_5;g_1,\ldots,g_7;\omega_1,\omega_2,\omega_3),\nonumber\\
(3,7,5,1) &=& (x_1,x_2,x_3; \psi_1,\ldots, \psi_7; g_1,\ldots,g_5; \omega).
\end{eqnarray}
Their respective supersymmetry transformations are 
\begin{eqnarray}
Q_1(1,5,7,3) &=& (\psi_4;g_5, -g_2, -g_3, {\dot x}, g_1;  {\dot \psi}_5,-{\dot \psi}_2, -{\dot \psi}_3,
\omega_3, {\dot \psi}_1, -\omega_1, -\omega_2; -{\dot g}_6, -{\dot g}_7, {\dot g}_4)\nonumber\\
Q_2(1,5,7,3) &=&  (\psi_1;{\dot x}, -g_7, g_6, -g_5, g_4;  -\omega_3, \omega_2, -\omega_1,
{\dot \psi}_5,-{\dot \psi}_4, {\dot \psi}_3,
-{\dot \psi}_2; -{\dot g}_3, {\dot g}_2, -{\dot g}_1)\nonumber\\
Q_3(1,5,7,3) &=&  (\psi_2; g_7,{\dot x}, -g_1, g_2, - g_3;  -{\dot \psi}_3,{\dot \psi}_4, -{\dot \psi}_5,
-\omega_1, \omega_2, -\omega_3,{\dot \psi}_1; -{\dot g}_4, {\dot g}_5, -{\dot g}_6)\nonumber\\
Q_4 (1,5,7,3) &=&  (\psi_3;-g_6, g_1,{\dot x}, g_3, g_2;  {\dot \psi}_2,{\dot \psi}_5, {\dot \psi}_4,
-\omega_2, -\omega_1, -{\dot \psi}_1, -\omega_3; -{\dot g}_5, -{\dot g}_4, -{\dot g}_7)\nonumber\\
Q_5(1,5,7,3) &=&  (\psi_5;-g_4, g_3, -g_2, -g_1, {\dot x}; - {\dot \psi}_4,-{\dot \psi}_3, {\dot \psi}_2,
-{\dot \psi}_1, \omega_3, \omega_2, -\omega_1; -{\dot g}_7, {\dot g}_6, {\dot g}_5)\nonumber
\end{eqnarray}
\begin{eqnarray}
&&
\end{eqnarray}
and
\begin{eqnarray}
Q_1(3,7,5,1) &=& (\psi_6,\psi_7,-\psi_4;-g_5, g_2, g_3, -{\dot x}_3, -g_1,   {\dot x}_1,{\dot x}_2; -{\dot \psi}_5,
{\dot \psi}_2,{\dot \psi}_3, -\omega, -{\dot\psi}_1;  -{\dot g}_4)\nonumber\\
Q_2(3,7,5,1) &=&  (\psi_3, -\psi_2,\psi_1;{\dot x}_3, -{\dot x}_2, {\dot x}_1, -g_5, g_4, -g_3, g_2; 
-\omega, {\dot \psi}_7,-{\dot \psi}_6, {\dot \psi}_5,
-{\dot \psi}_4;  -{\dot g}_1)\nonumber\\
Q_3(3,7,5,1) &=&  (\psi_4, -\psi_5, \psi_6;g_3, -g_4, g_5, {\dot x}_1, -{\dot x}_2, {\dot x}_3, -g_1;  
-{\dot \psi}_7,-\omega, {\dot \psi}_1, -{\dot \psi}_2,{\dot\psi}_3; -{\dot g}_2)\nonumber\\
Q_4 (3,7,5,1) &=&  (\psi_5, \psi_4, \psi_7;-g_2, -g_5, -g_4, {\dot x}_2, {\dot x}_1, g_1, {\dot x}_3;  {\dot \psi}_6,
-{\dot \psi}_1, -\omega, -{\dot \psi}_3,
-{\dot \psi}_2; -{\dot g}_3)\nonumber\\
Q_5(3,7,5,1) &=&  (\psi_7, -\psi_6, -\psi_5;g_4, g_3, -g_2, g_1, -{\dot x}_3, -{\dot x}_2,
{\dot x}_1;  {\dot \psi}_4,-{\dot \psi}_3, {\dot \psi}_2,
{\dot \psi}_1, -\omega; -{\dot g}_5)\nonumber
\end{eqnarray}
\begin{eqnarray}
&&
\end{eqnarray}
\par
{~}\par
{\em vii}) {\bf The quaternionic $N=5^{(\ast\ast)}$ irreps.}\par
They are recovered from the $Cl(6,1)$ Clifford algebra (see Section {\bf 3}).
This case admits $14$ inequivalent irreps. The length-$2$ irrep and the seven length-$3$
irreps are labeled as
\begin{eqnarray}
(8,8) &=& (x_1,\ldots x_8; \psi_1,\ldots, \psi_8),\nonumber\\
(7,8,1) &=& (x_1,\ldots ,x_7; \psi_1,\ldots, \psi_8;g),\nonumber\\
(6,8,2) &=& (x_1,\ldots,x_6; \psi_1,\ldots, \psi_8; g_1,g_2),\nonumber\\
(5,8,3) &=& (x_1,\ldots ,x_5; \psi_1,\ldots, \psi_8;g_1, g_2,g_3),\nonumber\\
(4,8,4) &=& (x_1,\ldots,x_4,; \psi_1,\ldots, \psi_8; g_1,\ldots , g_4),\nonumber\\
(3,8,5) &=& (x_1,x_2,x_3; \psi_1,\ldots, \psi_8;g_1,\ldots , g_5),\nonumber\\
(2,8,6) &=& (x_1,x_2; \psi_1,\ldots, \psi_8; g_1,\ldots ,g_6),\nonumber\\
(1,8,7) &=& (x; \psi_1,\ldots, \psi_8;g_1,\ldots, g_7),
\end{eqnarray}
the six extra length-$4$ irreps are given by\\
$(1,5,7,3), (3,7,5,1), (1,6,7,2), (2,7,6,1), (2,6,6,2), (1,7,7,1)$.\par
The length $2$ $(8,8)$ multiplet admits the following supersymmetry transformations
\begin{eqnarray}
Q_1(8,8) &=& (\psi_6, -\psi_5, -\psi_8, \psi_7, -\psi_2, \psi_1, \psi_4, -\psi_3;
{\dot x}_6, -{\dot x}_5, -{\dot x}_8,{\dot x}_7,-{\dot x}_2, {\dot x}_1, {\dot x}_4, -{\dot x}_3)\nonumber\\
Q_2(8,8) &=&  
(\psi_8, -\psi_7, \psi_6, -\psi_5, -\psi_4, \psi_3, -\psi_2, \psi_1;
{\dot x}_8, -{\dot x}_7, {\dot x}_6,-{\dot x}_5,-{\dot x}_4, {\dot x}_3, -{\dot x}_2, {\dot x}_1)
\nonumber\\
Q_3(8,8) &=&  
(\psi_7, \psi_8, -\psi_5, -\psi_6, -\psi_3, -\psi_4, \psi_1, \psi_2;
{\dot x}_7, {\dot x}_8, -{\dot x}_5,-{\dot x}_6,-{\dot x}_3,-{\dot x}_4, {\dot x}_1, {\dot x}_2)
\nonumber\\
Q_4 (8,8) &=& 
(\psi_5, \psi_6, \psi_7, \psi_8, \psi_1, \psi_2, \psi_3, \psi_4;
{\dot x}_5, {\dot x}_6, {\dot x}_7,{\dot x}_8,{\dot x}_1, {\dot x}_2, {\dot x}_3, {\dot x}_4)
\nonumber\\
Q_5(8,8) &=& 
(\psi_1, \psi_2, \psi_3, \psi_4, -\psi_5, -\psi_6, -\psi_7, -\psi_8;
{\dot x}_1, {\dot x}_2, {\dot x}_3,{\dot x}_4,-{\dot x}_5, -{\dot x}_6, -{\dot x}_7, -{\dot x}_8)\nonumber
\end{eqnarray}
\begin{eqnarray}
&&
\end{eqnarray}
The length $3$ irreps are immediately
read from the above transformations by setting, for any $p=1,\ldots, d-1$ associated with
the $(d-p,d,p)$ multiplet,
$g_1={\dot x}_8$, $g_2={\dot x}_7$, $\ldots$, $g_p={\dot x}_{9-p}$. The length-$4$ irreps are recovered from
similar transformations. To save
space the length $3$ and length $4$ irreps are not explicitly reported here.
\par
{~}\par
{\em viii}) {\bf The $N=6$ irreps}\par
The length $2$ and $3$ irreps are obtained from the $N=8$ irreps by restricting
the supersymmetry transformations to be given by $Q_i$, for $i=1,\ldots, 6$.
\par
This case admits $3$ extra, length-$4$, irreps which cannot be oxidized to $N=7$ and
an extra length-$4$ irrep oxidized to $N=7$. The three length-$4$ irreps with maximal
number of $N=6$ supersymmetries are given by
\begin{eqnarray}
(1,6,7,2) &=& (x; \psi_1,\ldots, \psi_6;g_1,\ldots,g_7;\omega_1,\omega_2),\nonumber\\
(2,7,6,1) &=& (x_1,x_2; \psi_1,\ldots, \psi_7;g_1,\ldots,g_6;\omega_1),\nonumber\\
(2,6,6,2) &=& (x_1,x_2; \psi_1,\ldots, \psi_6; g_1,\ldots,g_6; \omega_1, \omega_2),
\end{eqnarray}
The $(2,6,6,2)$ irrep is self-dual, while $(1,6,7,2)\leftrightarrow (2,7,6,1)$ are dually related
(see Section {\bf 4}). Their supersymmetry transformations are respectively given by 
\begin{eqnarray}
Q_1(1,6,7,2) &=& (\psi_1;{\dot x}, g_1, g_6, g_7,-g_4, -g_5;  {\dot \psi}_2,-\omega_1, -\omega_2,
-{\dot \psi}_5, -{\dot \psi}_6,
{\dot \psi}_3, {\dot\psi}_4; -{\dot g}_2,-{\dot g}_3)\nonumber\\
Q_2(1,6,7,2) &=&  (\psi_5;g_4, g_5, -g_2, -g_3, {\dot x}, g_1;  {\dot \psi}_6,-{\dot \psi}_3, -{\dot \psi}_4,
{\dot \psi}_1,{\dot\psi}_2, -\omega_1, -\omega_2; -{\dot g}_6, -{\dot g}_7)\nonumber\\
Q_3(1,6,7,2) &=&  (\psi_2;-g_1, {\dot x},-g_7, g_6, -g_5,g_4;  -{\dot \psi}_1,\omega_2, -\omega_1,
{\dot \psi}_6, -{\dot \psi}_5,
{\dot \psi}_4, -{\dot\psi}_3; -{\dot g}_3, {\dot g}_2)\nonumber\\
Q_4 (1,6,7,2) &=&  (\psi_3;-g_6, g_7,{\dot x}, -g_1, g_2,-g_3; - {\dot \psi}_4,{\dot \psi}_5, -{\dot \psi}_6,
-\omega_1,\omega_2,- {\dot \psi}_1,{\dot\psi}_2; -{\dot g}_4, {\dot g}_5)\nonumber\\
Q_5(1,6,7,2) &=&  (\psi_4;-g_7, -g_6, g_1, {\dot x}, g_3, g_2;  {\dot \psi}_3,{\dot \psi}_6, {\dot \psi}_5,
-\omega_2,-\omega_1, -{\dot \psi}_2, -{\dot\psi}_1; -{\dot g}_5, -{\dot g}_4)\nonumber\\
Q_6(1,6,7,2) &=&  (\psi_6;g_5, -g_4, g_3,-g_2,-g_1, {\dot x}; - {\dot \psi}_5,-{\dot \psi}_4, {\dot \psi}_3,
- {\dot \psi}_2, {\dot\psi}_1, \omega_2, -\omega_1; -{\dot g}_7, {\dot g}_6)\nonumber
\end{eqnarray}
\begin{eqnarray}&&
\end{eqnarray}
\begin{eqnarray}
Q_1(2,7,6,1) &=& (\psi_2,\psi_3;-g_2,{\dot x}_1,{\dot x}_2, g_5, g_6, -g_3, g_4; 
-\omega,- {\dot \psi}_1,-{\dot \psi}_6, -{\dot \psi}_7,
{\dot \psi}_4,{\dot\psi}_5; -{\dot g}_1)\nonumber\\
Q_2(2,7,6,1) &=&  (\psi_6,\psi_7;-g_6, g_3, g_4, -g_1, -g_2, {\dot x}_1,{\dot x}_2;  -{\dot \psi}_4,-{\dot \psi}_5, 
{\dot \psi}_2,{\dot \psi}_3, -\omega, -{\dot\psi}_1; -{\dot g}_5)\nonumber\\
Q_3(2,7,6,1) &=&  (\psi_3,-\psi_2;g_1,-{\dot x}_2, {\dot x}_1, -g_6, g_5, -g_4, g_3;  
{\dot \psi}_1,-\omega, {\dot \psi}_7, -{\dot \psi}_6,
{\dot \psi}_5, -{\dot\psi}_4; -{\dot g}_2)\nonumber\\
Q_4 (2,7,6,1) &=&  (\psi_4, -\psi_5;g_4, -g_5, g_6, {\dot x}_1,-{\dot x}_2, g_1, -g_2;  
{\dot \psi}_6,-{\dot \psi}_7, -\omega, {\dot \psi}_1,
-{\dot \psi}_2, {\dot\psi}_3; -{\dot g}_3)\nonumber\\
Q_5(2,7,6,1) &=&  (\psi_5, \psi_4;-g_3, -g_6, -g_5, {\dot x}_2,{\dot x}_1, g_2, g_1;  
{\dot \psi}_7,{\dot \psi}_6, -{\dot \psi}_1,
-\omega, -{\dot \psi}_3, -{\dot\psi}_2; -{\dot g}_4)\nonumber\\
Q_6(2,7,6,1) &=&  (\psi_7, -\psi_6;g_5, g_4, -g_3,g_2,-g_1, {\dot x}_2, {\dot x}_1;  
-{\dot \psi}_5,{\dot \psi}_4, -{\dot \psi}_3,
{\dot \psi}_2, {\dot\psi}_1,-\omega; -{\dot g}_6)\nonumber
\end{eqnarray}
\begin{eqnarray}
&&\end{eqnarray}
and
\begin{eqnarray}
Q_1(2,6,6,2) &=& (\psi_1, \psi_2;{\dot x}_1, {\dot x}_2, g_5, g_6, -g_3, - g_4;  
-\omega_1,-\omega_2, -{\dot \psi}_5,-{\dot \psi}_6, {\dot \psi}_3,
{\dot \psi}_4; -{\dot g}_1, -{\dot g}_2)\nonumber\\
Q_2(2,6,6,2) &=&  (\psi_5, \psi_6;g_3, g_4, -g_1, -g_2, {\dot x}_1,{\dot x}_2;  
-{\dot \psi}_3,-{\dot \psi}_4, {\dot \psi}_1,
{\dot \psi}_2, -\omega_1, -\omega_2; -{\dot g}_5, -{\dot g}_6)\nonumber\\
Q_3(2,6,6,2) &=&  (\psi_2, -\psi_1;-{\dot x}_2, {\dot x}_1, -g_6, g_5, -g_4, g_3;  
\omega_2, -\omega_1, {\dot \psi}_6,-{\dot \psi}_5, {\dot \psi}_4,
-{\dot \psi}_3; -{\dot g}_2, {\dot g}_1)\nonumber\\
Q_4 (2,6,6,2) &=&  (\psi_3, -\psi_4;-g_5, g_6, {\dot x}_1, -{\dot x}_2, g_1, -g_2;  
{\dot \psi}_5,-{\dot \psi}_6, -\omega_1, \omega_2, -{\dot \psi}_1,
{\dot \psi}_2; -{\dot g}_3, {\dot g}_4)\nonumber\\
Q_5(2,6,6,2) &=&  (\psi_4, \psi_3;-g_6, -g_5, {\dot x}_2, {\dot x}_1, g_2, g_1;  
{\dot \psi}_6,{\dot \psi}_5, -\omega_2, -\omega_1, -{\dot \psi}_2,
-{\dot \psi}_1; -{\dot g}_4, -{\dot g}_3)\nonumber\\
Q_6(2,6,6,2) &=&  (\psi_6, -\psi_5;g_4, -g_3, g_2, -g_1, -{\dot x}_2, {\dot x}_1;  
-{\dot \psi}_4,{\dot \psi}_3, -{\dot \psi}_2,
{\dot \psi}_1, \omega_2, -\omega_1; -{\dot g}_6, {\dot g}_5)\nonumber
\end{eqnarray}
\begin{eqnarray}&&
\end{eqnarray}
\par
{~}\par
{\em ix}) {\bf The $N=7$ irreps}\par
The length $2$ and $3$ irreps are obtained from the $N=8$ irreps by restricting
the supersymmetry transformations to be given by $Q_i$, for $i=1,\ldots, 7$.
\par
This case admits an extra, length-$4$, irrep
\begin{eqnarray}
(1,7,7,1) &=& (x; \psi_1,\ldots, \psi_7;g_1,\ldots,g_7;\omega),
\end{eqnarray}
whose supersymmetry transformations are
\begin{eqnarray}
Q_1(1,7,7,1) &=& (\psi_2;-g_3,{\dot x}, g_1, g_6, g_7, -g_4, -g_5;  
{\dot \psi}_3,-\omega, -{\dot \psi}_1, -{\dot \psi}_6,
-{\dot \psi}_7,{\dot\psi}_4, {\dot\psi}_5; -{\dot g}_2)\nonumber\\
Q_2(1,7,7,1) &=&  (\psi_6;-g_7, g_4, g_5, -g_2, -g_3, {\dot x}, g_1;  
{\dot \psi}_7,-{\dot \psi}_4, -{\dot \psi}_5,
{\dot \psi}_2,{\dot\psi}_3,  -\omega, -{\dot\psi}_1; -{\dot g}_6)\nonumber\\
Q_3(1,7,7,1) &=&  (\psi_1;{\dot x}, g_3, -g_2, -g_5, g_4, g_7, -g_6;  
-\omega, -{\dot \psi}_3,{\dot \psi}_2, {\dot \psi}_5,
-{\dot \psi}_4, -{\dot\psi}_7, {\dot\psi}_6; -{\dot g}_1)\nonumber\\
Q_4 (1,7,7,1) &=&  (\psi_3;
g_2, -g_1,{\dot x}, -g_7, g_6, -g_5, g_4;  
-{\dot \psi}_2,{\dot \psi}_1, -\omega, {\dot \psi}_7,
-{\dot \psi}_6,{\dot\psi}_5, -{\dot\psi}_4; -{\dot g}_3)\nonumber\\
Q_5(1,7,7,1) &=&  (\psi_4;
g_5, -g_6, g_7, {\dot x}, -g_1, g_2, -g_3;  
-{\dot \psi}_5,{\dot \psi}_6, -{\dot \psi}_7,
-\omega, {\dot \psi}_1, -{\dot\psi}_2, {\dot\psi}_3; -{\dot g}_4)\nonumber\\
Q_6(1,7,7,1) &=&  (\psi_5;-g_4, -g_7, -g_6, g_1, {\dot x}, g_3, g_2;  
{\dot \psi}_4,{\dot \psi}_7, {\dot \psi}_6,
- {\dot \psi}_1, -\omega, -{\dot\psi}_3, -{\dot\psi}_2; -{\dot g}_5)\nonumber\\
Q_7 (1,7,7,1) &=&  (\psi_7;g_6, g_5, -g_4, g_3, -g_2, -g_1, {\dot x};  
-{\dot \psi}_6,-{\dot \psi}_5, {\dot \psi}_4,
-{\dot \psi}_3, {\dot\psi}_2, {\dot\psi}_1, -\omega; -{\dot g}_7)\nonumber
\end{eqnarray}
\begin{eqnarray}
&&
\end{eqnarray}
\par
{~}\par
{\em x}) {\bf The $N=8$ irreps}\par
The $8$ inequivalent irreps are here listed
\begin{eqnarray}
(8,8) &=& (x_1,\ldots x_8; \psi_1,\ldots, \psi_8),\nonumber\\
(7,8,1) &=& (x_1,\ldots ,x_7; \psi_1,\ldots, \psi_8;g),\nonumber\\
(6,8,2) &=& (x_1,\ldots,x_6; \psi_1,\ldots, \psi_8; g_1,g_2),\nonumber\\
(5,8,3) &=& (x_1,\ldots ,x_5; \psi_1,\ldots, \psi_8;g_1, g_2,g_3),\nonumber\\
(4,8,4) &=& (x_1,\ldots,x_4,; \psi_1,\ldots, \psi_8; g_1,\ldots , g_4),\nonumber\\
(3,8,5) &=& (x_1,x_2,x_3; \psi_1,\ldots, \psi_8;g_1,\ldots , g_5),\nonumber\\
(2,8,6) &=& (x_1,x_2; \psi_1,\ldots, \psi_8; g_1,\ldots ,g_6),\nonumber\\
(1,8,7) &=& (x; \psi_1,\ldots, \psi_8;g_1,\ldots, g_7).
\end{eqnarray}
The length-$2$ $(8,8)$ multiplet admits the following supersymmetry transformations
\begin{eqnarray}
Q_1(8,8) &=&
(\psi_3, \psi_4, -\psi_1, -\psi_2, -\psi_7, -\psi_8, \psi_5, \psi_6;
-{\dot x}_3, -{\dot x}_4, {\dot x}_1,{\dot x}_2,{\dot x}_7, {\dot x}_8, -{\dot x}_5, -{\dot x}_6)
\nonumber\\
Q_2(8,8) &=&  
(\psi_7, \psi_8, -\psi_5, -\psi_6, \psi_3, \psi_4, -\psi_1, -\psi_2;
-{\dot x}_7, -{\dot x}_8, {\dot x}_5,{\dot x}_6,-{\dot x}_3, -{\dot x}_4, {\dot x}_1, {\dot x}_2)
\nonumber\\
Q_3(8,8) &=& 
(\psi_2, -\psi_1, -\psi_4, \psi_3, \psi_6, -\psi_5, -\psi_8, \psi_7;
-{\dot x}_2, {\dot x}_1, {\dot x}_4,-{\dot x}_3,-{\dot x}_6, {\dot x}_5, {\dot x}_8, -{\dot x}_7)
\nonumber\\
Q_4 (8,8) &=& 
(\psi_4, -\psi_3, \psi_2, -\psi_1, \psi_8, -\psi_7, \psi_6, -\psi_5;
-{\dot x}_4, {\dot x}_3, -{\dot x}_2,{\dot x}_1,-{\dot x}_8, {\dot x}_7, -{\dot x}_6, {\dot x}_5)
\nonumber\\
Q_5(8,8) &=&  
(\psi_5, -\psi_6, \psi_7, -\psi_8, -\psi_1, \psi_2, -\psi_3, \psi_4;
-{\dot x}_5, {\dot x}_6, -{\dot x}_7,{\dot x}_8,{\dot x}_1, -{\dot x}_2, {\dot x}_3, -{\dot x}_4)
\nonumber\\
Q_6(8,8) &=&  
(\psi_6, \psi_5, \psi_8, \psi_7, -\psi_2, -\psi_1, -\psi_4, -\psi_3;
-{\dot x}_6, -{\dot x}_5, -{\dot x}_8,-{\dot x}_7,{\dot x}_2, {\dot x}_1, {\dot x}_4, {\dot x}_3)
\nonumber\\
Q_7 (8,8) &=&  
(\psi_8, -\psi_7, -\psi_6, \psi_5, -\psi_4, \psi_3, \psi_2, -\psi_1;
-{\dot x}_8, {\dot x}_7, {\dot x}_6,-{\dot x}_5,{\dot x}_4, -{\dot x}_3, -{\dot x}_2, {\dot x}_1)
\nonumber\\
Q_8(8,8) &=&  
(\psi_1, \psi_2, \psi_3, \psi_4, \psi_5, \psi_6, \psi_7, \psi_8;
{\dot x}_1, {\dot x}_2, {\dot x}_3,{\dot x}_4,{\dot x}_5, {\dot x}_6, {\dot x}_7, {\dot x}_8)
\end{eqnarray}
The length $3$ irreps are immediately
read from the above transformations by setting, for any $p=1,\ldots, d-1$ associated with
the $(d-p,d,p)$ multiplet,
$g_1={\dot x}_8$, $g_2={\dot x}_7$, $\ldots$, $g_p={\dot x}_{9-p}$. To save
space the length $3$ irreps are not explicitly reported here.
\par
{~}\par
\renewcommand{\theequation}{B.\arabic{equation}}
\setcounter{equation}{0}
 
{\Large{\bf Appendix B\\{~}\\ Classification of the $N=9,10$ irreps
and length-$4$ 
$N=11^{(\ast)}, 12$ irreps}}\par
{~}\par
We present here the complete classification of irreps of the 
$N=9,10$ supersymmetries producing (the length $l=2$ and $l=3$ irreps being already known) the whole list
of length $l\geq 4$ inequivalent irreps. $N=9$ does not admit length $l\geq 5$ irreps, while $N=10$ is the
lowest number of extended supersmmetries admitting irreps with length $l>4$. For $N=10$ the maximal 
length $L$ of its irreps is $L=5$. \par
We further produce the list of length-$4$ inequivalent irreps for the next two values of the
oxidized supersymmetries, namely
$N=11^{(\ast)}$ and $N=12$.\par
{~}\par
{\em i}) {\bf Classification of the $N=9$ irreps}\par
{~}\par
The length-$4$ irreducible multiplet $(d_1,d_2,d_3,d_4)$ is for simplicity expressed in terms of the two
positive integers $h\equiv d_1$, $k=d_4$, since $d_3= 16-h$, $d_2=16-k$.\par
$N=9$ presents $4$ length-$4$ irreducible self-dual (under (\ref{hiloduality})) multiplets for
\begin{eqnarray}
&h=k=1,2,3,4&
\end{eqnarray}
and $2\times(6+4+2)= 24$ non self-dual length-$4$ irreducible multiplets given by
the series of coupled values
\begin{eqnarray}
h=1 &\& & k=2,\ldots ,7\nonumber\\
h=2 &\& & k=3,\ldots, 6\nonumber\\
h=3 &\& & k=4,5
\end{eqnarray}
together with the $(h\leftrightarrow k)$ dually interchanged multiplets.\par
The previous results can be summarized as follows. Inequivalent length-$4$ irreps are in $1$-to-$1$
correspondence with the ordered pair of positive integers $h,k$
satisfying the constraint
\begin{eqnarray}
h+k &\leq& 8. 
\end{eqnarray} 
The total number ${\overline k}_4$
of inequivalent length-$4$ irreps (without discriminating, see (\ref{kappa}), the statistics of the multiplets) is
\begin{eqnarray}
{\overline k}_4 &=& 28
\end{eqnarray}
\par
{~}\par
{\em ii}) {\bf Classification of the $N=10$ irreps}\par
{~}\par
$N=10$ admits irreps up to length $l=5$. We have\par
{~}\par
{\em iia}) {\em The length-$4$ classification.}\par
{~}\par
The length-$4$ irreducible multiplet $(d_1,d_2,d_3,d_4)$ is for simplicity expressed in terms of the two
positive integers $h\equiv d_1$, $k=d_4$, since $d_3= 32-h$, $d_2=32-k$.\par
$N=10$ presents $8$ length-$4$ irreducible self-dual (under (\ref{hiloduality})) multiplets for
\begin{eqnarray}
&h=k=1,2,\ldots, 8&
\end{eqnarray}
and a set of 
$2\times 3(\sum_{j=1}^{7}j)= 168$ non self-dual length-$4$ irreducible multiplets given by
the series of coupled values
\begin{eqnarray}
h=1 &\& & k=2,\ldots ,22\nonumber\\
h=2 &\& & k=3,\ldots, 20\nonumber\\
h=3 &\& & k=4,\ldots ,18\nonumber\\
h=4 &\& & k=5,\ldots, 16\nonumber\\
h=5 &\& & k=6,\ldots ,14\nonumber\\
h=6 &\& & k=7,\ldots, 12\nonumber\\
h=7 &\& & k=8,9,10
\end{eqnarray}
together with the $(h\leftrightarrow k)$ dually interchanged multiplets.\par
If we set 
\begin{eqnarray}\label{minim}
r &=& min(h,k)
\end{eqnarray}
the previous results can be summarized as follows. Inequivalent length-$4$ irreps are in $1$-to-$1$
correspondence with the ordered pair of positive integers $h,k$
satisfying the constraint
\begin{eqnarray}
h+k+r & \leq & 24.
\end{eqnarray}
The total number ${\overline k}_4$
of inequivalent length-$4$ irreps (without discriminating, see (\ref{kappa}), the statistics of the multiplets) is
\begin{eqnarray}
{\overline k}_4 &=& 176
\end{eqnarray}
\par
{~}\par

{\em iib}) {\em The length-$5$ classification}\par
{~}\par
A length-$5$ multiplet 
$(d_1,d_2,d_3,d_4,d_5)$ is characterized by three independent positive integers, let's say 
$d_1,d_2,d_5$, since
$d_4= 32-d_2$ and $d_3=32-d_1-d_5$.
The full list of length-$5$ irreps of the $N=10$ supersymmetry can be listed
according to the number $d_5$ of highest-spin auxiliary fields.
The maximal number of auxiliary fields is $7$. At any fixed $d_5=1,\ldots ,7$ the number
of inequivalent irreps is $(8-{d_5})^2$. Therefore, the total number
${\overline k}_5$ of length-$5$ inequivalent irreps is given by 
\begin{eqnarray}
&{\overline k}_5= 1^2+2^2+\ldots + 7^2=140&
\end{eqnarray}
The full list of irreps is here produced in terms, at any fixed $d_5$, of the ordered ${\underline{ d_1,d_2}}$ pairs.
We have
\begin{eqnarray}
d_5=7 &:& {\underline{1,10}}.\nonumber\\
d_5=6 &:& {\underline{1,10}},{\underline{1,11}},{\underline{1,12}}/{\underline{2,12}}.\nonumber\\
d_5=5 &:&  {\underline{1,10}}, \ldots ,{\underline{1,14}}/
{\underline{2,12}}, \ldots ,{\underline{2,14}}/{\underline{3,14}}. \nonumber\\
d_5=4&:& {\underline{1,10}}, \ldots ,{\underline{1,16}}/
{\underline{2,12}}, \ldots ,{\underline{2,16}}/ {\underline{3,14}}, \ldots,{\underline{3,16}}/
{\underline{4,16}}. \nonumber\\
d_5=3&:& {\underline{1,10}}, \ldots ,{\underline{1,18}}/
{\underline{2,12}}, \ldots ,{\underline{2,18}}/ {\underline{3,14}}, \ldots,{\underline{3,18}}/
{\underline{4,16}}, \ldots,{\underline{4,18}}/{\underline{5,18}} . \nonumber\\
d_5=2&:& {\underline{1,10}}, \ldots ,{\underline{1,20}}/
{\underline{2,12}}, \ldots ,{\underline{2,20}}/ {\underline{3,14}}, \ldots,{\underline{3,20}}/
{\underline{4,16}}, \ldots,{\underline{4,20}}/\nonumber\\
&& {\underline{5,18}}, \ldots , {\underline{5,20}}/{\underline{6,20}}.\nonumber\\
d_5=1&:& {\underline{1,10}}, \ldots ,{\underline{1,22}}/
{\underline{2,12}}, \ldots ,{\underline{2,22}}/ {\underline{3,14}}, \ldots,{\underline{3,22}}/
{\underline{4,16}}, \ldots,{\underline{4,22}}/\nonumber\\
&& {\underline{5,18}}, \ldots , {\underline{5,22}}/{\underline{6,20}},\ldots, 
{\underline{6,22}}/{\underline{7,22}}.
\end{eqnarray}
One can check that the above set of irreducible multiplets is indeed closed under the
(\ref{hiloduality}) {\em high $\Leftrightarrow$ low spin} duality transformations.
\par
{~}\par
{\em iii}) {\bf Classification of the length-$4$ $N=11^{(\ast)}$ irreps}\par
{~}\par
The length-$4$ irreducible multiplet $(d_1,d_2,d_3,d_4)$ is for simplicity expressed in terms of the two
positive integers $h\equiv d_1$, $k=d_4$, since $d_3= 64-h$, $d_2=64-k$.\par
$N=11^{(\ast)}$ presents $16$ length-$4$ irreducible self-dual (under (\ref{hiloduality})) multiplets for
\begin{eqnarray}
&h=k=1,2,\ldots , 16&
\end{eqnarray}
and $776$ non self-dual length-$4$ irreducible multiplets given by
the series of coupled values
\begin{eqnarray}&\begin{tabular}{ll}
$h=1 ~\&~  k=2,\ldots ,53$~\quad&\quad $h=9~ ~\&~k=10,\ldots ,30$\\
$h=2 ~\&~  k=3,\ldots, 50$\quad&\quad$h=10 ~\& ~ k=11,\ldots ,28$\\
$h=3 ~\&~  k=4,\ldots ,47$\quad&\quad$h=11 ~\& ~ k=12,\ldots ,26$\\
$h=4 ~\&~  k=5,\ldots, 44$\quad&\quad$h=12 ~\& ~ k=13,\ldots ,24$\\
$h=5 ~\&~  k=6,\ldots ,41$\quad&\quad$h=13 ~\& ~ k=14,\ldots ,22$\\
$h=6 ~\&~  k=7,\ldots, 38$\quad&\quad$h=14 ~\& ~ k=15,\ldots ,20$\\
$h=7 ~\&~  k=8,\ldots, 35$\quad&\quad$h=15 ~\& ~ k=16,\ldots ,18$\\
$h=8 ~\&~  k=9,\ldots, 32$\quad &
\end{tabular}
\end{eqnarray}
together with the $(h\leftrightarrow k)$ dually interchanged multiplets.\par
The previous results can be summarized as follows. 
Let us set, as before (\ref{minim}), $r= min (h,k)$ and introduce the $s(r)$ function defined through
\begin{eqnarray}
s(r) &=& \left\{
\begin{tabular}{cl} 
$8-r$ \quad & for $r=1,\ldots , 7$\\
$0$ \quad & otherwise
\end{tabular}
\right\}
\end{eqnarray}   
Inequivalent length-$4$ irreps are in $1$-to-$1$
correspondence with the ordered pair of positive integers $h,k$
satisfying the constraint
\begin{eqnarray}
h+k+r-s(r) &\leq& 48.
\end{eqnarray}
The total number ${\overline k}_4$
of inequivalent length-$4$ irreps (without discriminating, see (\ref{kappa}), the statistics of the multiplets) is
\begin{eqnarray}
{\overline k}_4 &=& 792.
\end{eqnarray}
\par
{~}\par

{\em iii}) {\bf Classification of the length-$4$ $N=12$ irreps}\par
{~}\par
The length-$4$ irreducible multiplet $(d_1,d_2,d_3,d_4)$ is for simplicity expressed in terms of the two
positive integers $h\equiv d_1$, $k=d_4$, since $d_3= 64-h$, $d_2=64-k$.\par
$N=12$ presents $12$ length-$4$ irreducible self-dual (under (\ref{hiloduality})) multiplets for
\begin{eqnarray}
&h=k=1,2,\ldots , 12&
\end{eqnarray}
and $584$ non self-dual length-$4$ irreducible multiplets given by
the series of coupled values
\begin{eqnarray}&\begin{tabular}{ll}
$h=1 ~\&~  k=2,\ldots ,52$~\quad&\quad $h=7~ ~\&~k=~8,\ldots ,28$\\
$h=2 ~\&~  k=3,\ldots, 48$\quad&\quad$h=8~ ~\& ~ k=~9,\ldots ,24$\\
$h=3 ~\&~  k=4,\ldots ,44$\quad&\quad$h=9~ ~\& ~ k=10,\ldots ,21$\\
$h=4 ~\&~  k=5,\ldots, 40$\quad&\quad$h=10 ~\& ~ k=11,\ldots ,18$\\
$h=5 ~\&~  k=6,\ldots ,36$\quad&\quad$h=11 ~\& ~ k=12,\ldots ,15$\\
$h=6 ~\&~  k=7,\ldots, 32$\quad&\\
\end{tabular}
&
\end{eqnarray}
together with the $(h\leftrightarrow k)$ dually interchanged multiplets.\par
The total number ${\overline k}_4$
of inequivalent length-$4$ irreps (without discriminating, see (\ref{kappa}), the statistics of the multiplets) is
\begin{eqnarray}
{\overline k}_4 &=& 596.
\end{eqnarray}
\par
{~}\par

\renewcommand{\theequation}{C.\arabic{equation}}
\setcounter{equation}{0}
\par
{~}\par
{\Large{\bf Appendix C\\{~}\\ Irreps decompositions of multiplets tensor products}}\par
{~}\par
As discussed in Section {\bf 6}, multilinear terms entering a manifestly invariant 
$N$-extended supersymmetric action can be extracted by (multiple) tensor products of irreps.
In our framework this procedure replaces the supersymmetric calculus. The advantage of our method
consists in the fact that it can be systematically carried out for any arbitrary value of $N$
(the limitations are only due to the increasing computational complexity and are not of conceptual nature), 
even when the superfield formalism is not available. 
For illustrative purposes it is convenient to present here some explicit examples which have been
discussed in the main text. 
\par
{~}\par
{\em i}) {\bf Tensoring the $N=2$ bosonic irreps}\par
{~}\par
The two inequivalent $N=2$ bosonic irreps, $(1,2,1)$ and $(2,2)$, correspond in the superfield language
to, respectively, the real and chiral $N=2$ linear bosonic superfields. In our framework we get
the following results, for their mutual tensoring:  
\par
{~}\par
{\em ia}) {\em $(1,2,1)_{s=0}\times (1,2,1)_{s=0}= (1,2,1)_{\parallel s=0} +(\ldots)_{s>0}$}\par
{~}\par
${(\ldots)}_{s>0}$ on the r.h.s denotes the non-leading (i.e. of higher spin) bilinear irreps.  
The suffix ``${}_\parallel$" on the $(1,2,1)_{\parallel s=0}$ multiplet on the r.h.s. means that it is
{\em symmetric}, i.e. it is non-vanishing when the left and right multiplets in the l.h.s. are identified.
Let us set $(x;\psi_1,\psi_2;g)$, $(y;\lambda_1,\lambda_2;f)$ the two multiplets on the l.h.s. . The bilinear
multiplet ${(1,2,1)_{\parallel s=0}} \equiv
({\tilde x}; {{\tilde \psi}_1}, {{\tilde \psi}_2}; {\tilde g})$ on the r.h.s. is therefore given by
\begin{eqnarray}
{\tilde x}&=& xy\nonumber\\
{{\tilde \psi}_1} &=&\psi_1y +x\lambda_1\nonumber\\
{{\tilde \psi}_2} &=&\psi_2y+x\lambda_2\nonumber\\
{\tilde g} &=&gy -\psi_1\lambda_2+\psi_2\lambda_1+xf
\end{eqnarray}
\par
{~}\par
{\em ib}) {\em $(1,2,1)_{s=0}\times (2,2)_{s=0}= 2\times (1,2,1)_{s=0} + (\ldots)_{s>0}$}\par
{~}\par
The second multiplet on the l.h.s. is here given by $(y_1,y_2;\lambda_1,\lambda_2)$. The two leading
bilinear multiplets in the r.h.s. are ${(1,2,1)_{\parallel s=0}}^{a,b} \equiv
({\tilde x}^{a,b}; {{\tilde \psi}_1}^{a,b}, {{\tilde \psi}_2}^{a,b}; {\tilde g}^{a,b})$,
with
\begin{eqnarray}
{\tilde x}^a &=&x_1y\nonumber\\
{{\tilde \psi}_1}^a &=&\psi_1y+x_1\lambda_1\nonumber\\
{{\tilde \psi}_2}^a &=&\psi_2y+x_1\lambda_2\nonumber\\
{\tilde g}^a &=&{\dot x}_2y -\psi_1\lambda_2+\psi_2\lambda_1+x_1f
\end{eqnarray}
and
\begin{eqnarray}
{\tilde x}^b &=&x_2y\nonumber\\
{{\tilde \psi}_1}^b &=&-\psi_2y +x_2\lambda_1\nonumber\\
{{\tilde \psi}_2}^b &=&\psi_1y +x_2\lambda_2\nonumber\\
{\tilde g}^b &=&-{\dot x}_1y +\psi_2\lambda_2+\psi_1\lambda_1+x_2f
\end{eqnarray}
 \par{~}\par
{\em ic}) {\em  $(2,2)_{s=0}\times (2,2)_{s=0} = (2,2)_{\parallel s=0}+ (1,2,1)_{\parallel s=0} + (1,2,1)_{\perp s=0} +
(\ldots )_{s>0}$}\par
{~}\par
Tensoring two length-$2$ multiplets, $(x_1,x_2;\psi_1,\psi_2)$, $(y_1,y_2;\lambda_1,\lambda_2)$,
produces a bilinear {\em antisymmetric} leading ($s=0$) multiplet $(1,2,1)_{\perp s=0} \equiv ({\overline x}, {\overline \psi}_1, {\overline\psi}_2, {\overline g})$
which vanishes when 
$(x_1,x_2;\psi_1,\psi_2)\equiv (y_1,y_2;\lambda_1,\lambda_2)$ are 
identified, 
plus two {\em symmetric} leading bilinear multiplets, given
by
$(2,2)_{\parallel s=0} \equiv ({\tilde x}_1, {\tilde x}_2; {\tilde \psi}_1, {\tilde\psi}_2)$,
$(1,2,1)_{\parallel s=0} \equiv ({\hat x}; {\hat \psi}_1, {\hat\psi}_2; {\hat g})$.\par
We have, explicitly,
\begin{eqnarray}
{\overline x} &=&x_1y_2-x_2y_1\nonumber\\
{\overline \psi}_1 &=&\psi_1y_2-x_1\lambda_2+\psi_2y_1-x_2\lambda_1\nonumber\\
{\overline \psi}_2 &=&\psi_2y_2+x_1\lambda_1-\psi_1y_1-x_2\lambda_2\nonumber\\
{\overline g} &=& -2\psi_1\lambda_1-2\psi_2\lambda_2+{\dot x}_2y_2-x_2{\dot y}_2+{\dot x}_1y_1-x_1{\dot y}_1,
\end{eqnarray}
\begin{eqnarray}
{\tilde x}_1 &=&x_1y_1-x_2y_2\nonumber\\
{\tilde x}_2 &=&x_1y_2+x_2y_1\nonumber\\
{\tilde \psi}_1 &=&\psi_1y_1+x_1\lambda_1+\psi_2y_2+x_2\lambda_2\nonumber\\
{\tilde \psi}_2 &=&\psi_2y_1+x_1\lambda_2-\psi_1y_2-x_2\lambda_1,
\end{eqnarray}
and
\begin{eqnarray}
{\hat x} &=&x_1y_1+x_2y_2\nonumber\\
{\hat \psi}_1 &=&\psi_1y_1-\psi_2y_2+x_1\lambda_1-x_2\lambda_2\nonumber\\
{\hat \psi}_2 &=&\psi_2y_1+\psi_1y_2+x_1\lambda_2+x_2\lambda_1\nonumber\\
{\hat g} &=&2\psi_2\lambda_1-2\psi_1\lambda_2+{\dot x}_2y_1-{\dot x}_1y_2+x_1{\dot y}_2-x_2{\dot y}_1.
\end{eqnarray}
\par
{~}\par
{\em ii}) {\bf Tensoring the $N=4$ bosonic irreps (selected cases)}\par
{~}\par
We present here the mutual tensoring of the $(1,4,3)$ and the $(2,4,2)$ irreps.
\par
{~}\par
{\em iia}) {\em The $(1,4,3)_{s=0}\times (1,4,3)_{s=0}$ case}\par
{~}\par
Let us denote the left multiplet as $(x; \psi_1,\psi_2,\psi_3,\psi_4;g_1,g_2,g_3)$ and the right 
multiplet as $(y; \lambda_1,\lambda_2,\lambda_3,\lambda_4;f_1,f_2,f_3)$. Their tensor product gives at the 
leading order ($s=0$) the (reducible) enveloping representation
of the $N=4$ supersymmetry, given by the $(1,4,6,4,1)$ multiplet with elements
${\bf 1}$, $Q_i{\bf 1}$ for $i=1,2,3,4$, $Q_iQ_j{\bf 1}$ for $i<j$, $Q_iQ_jQ_k{\bf 1}$
for $i<j<k$ and, finally, $Q_1Q_2Q_3Q_4{\bf 1}$. \par
The following identifications hold
\begin{eqnarray}
{\bf 1} &=& xy,\nonumber\\
Q_1 {\bf 1}&=& \psi_2y+x\lambda_2\nonumber\\
Q_2{\bf 1}&=&\psi_4 y + x\lambda_4\nonumber\\
Q_3{\bf 1}&=&\psi_3 y + x\lambda_3\nonumber\\
Q_4{\bf 1}&=&\psi_1 y + x\lambda_1\nonumber\\
Q_1Q_2{\bf 1}&=&\psi_2\lambda_4-g_2 y -\psi_4\lambda_2-xf_2\nonumber\\
Q_1Q_3{\bf 1}&=&\psi_2\lambda_3+g_3y-\psi_3\lambda_2+xf_3\nonumber\\
Q_1Q_4{\bf 1}&=&\psi_2\lambda_1-g_1y-\psi_1\lambda_2-xf_1\nonumber\\
Q_2Q_3{\bf 1}&=&\psi_4\lambda_3-g_1y-\psi_3\lambda_4-xf_1\nonumber\\
Q_2Q_4{\bf 1}&=&\psi_4\lambda_1-g_3y-\psi_1\lambda_4-xf_3\nonumber\\
Q_3Q_4{\bf 1}&=&\psi_3\lambda_1-g_2y-\psi_1\lambda_3-xf_2\nonumber\\
Q_1Q_2Q_3{\bf 1}&=&{\dot\psi}_1y-g_2\lambda_3-g_3\lambda_4-{\psi_2}f_1-g_1\lambda_2-\psi_4f_3-\psi_3f_2+
x{\dot\lambda}_1\nonumber\\
Q_1Q_2Q_4{\bf 1}&=&-{\dot\psi}_3y-g_2\lambda_1+g_1\lambda_4-\psi_2f_3-g_3\lambda_2+\psi_4f_1-\psi_1f_2-
x{\dot\lambda}_3\nonumber\\
Q_1Q_3Q_4{\bf 1}&=&{\dot\psi}_4y+g_3\lambda_1+g_1\lambda_3-\psi_2f_2-g_2\lambda_2+\psi_3f_1+\psi_1f_3+
x{\dot\lambda}_4\nonumber\\
Q_2Q_3Q_4{\bf 1}&=&-{\dot\psi}_2y-g_1\lambda_1+g_3\lambda_3-\psi_4f_2-g_2\lambda_4+\psi_3f_3-\psi_1f_1-
x{\dot\lambda}_2\nonumber\\
Q_1Q_2Q_3Q_4{\bf 1}&=& -{\ddot x}y-x{\ddot y}+{\dot\psi}_1\lambda_1+{\dot\psi}_2\lambda_2+{\dot\psi}_3\lambda_3+
{\dot\psi}_4\lambda_4-\psi_1{\dot\lambda}_1-\psi_2{\dot\lambda}_2-\nonumber\\&&
-\psi_3{\dot\lambda}_3-\psi_4
{\dot\lambda}_4+2g_1f_1+2g_2f_2+2g_3f_3
\end{eqnarray}
Since the identity ${\bf 1}$ is bosonic, therefore  $\Gamma^5{\bf 1} ={\bf 1}$, see (\ref{fermionnumber}).\par
$Q_1Q_2{\bf 1}-Q_3Q_4{\bf 1}$, $Q_2Q_3{\bf 1}-Q_1Q_4{\bf 1}$, $Q_3Q_1{\bf 1}-Q_2Q_4{\bf 1}$ are the three leading
bosonic fields of a $(3,4,1)_{s=1}$ irrep contained in the enveloping representation as a subrepresentation. Quotienting
out such an irrep from the enveloping representation (by consistently
setting all eight corresponding fields identically equal to zero, see (\ref{adjointdecomp})) we
obtain the $(1,4,3)_{s=0}$ irrep, in terms of a single spin $0$ field (the identity ${\bf 1}$), four spin $\frac{1}{2}$
fields (given by $Q_i{\bf 1}$) and three spin $1$ fields ($Q_1Q_2{\bf 1}+Q_3Q_4{\bf 1}$, $Q_2Q_3{\bf 1}+Q_1Q_4{\bf 1}$, $Q_3Q_1{\bf 1}+Q_2Q_4{\bf 1}$) which play the role of auxiliary fields.\par
{~}\par

{\em iib}) {\em The $(2,4,2)_{s=0}\times (1,4,3)_{s=0}$ case} \par
{~}\par
Let us denote now as $(x_1,x_2; \psi_1,\psi_2,\psi_3,\psi_4;g_1,g_2)$ the left multiplet
and the right multiplet as $(y; \lambda_1,\lambda_2,\lambda_3,\lambda_4;f_1,f_2,f_3)$. Their tensor product 
produces at the leading ($s=0$) order two (reducible) enveloping representations
of the $N=4$ supersymmetry, given by two $(1,4,6,4,1)$ multiplets.
They are respectively lead by ${\bf 1}\equiv x_1 y$ and by ${\bf 1}\equiv x_2 y$.
Explicitly, we get in the first case
\begin{eqnarray}
{\bf 1} &=& x_1y,\nonumber\\
Q_1 {\bf 1}&=& \psi_2y+x_1\lambda_2\nonumber\\
Q_2{\bf 1}&=&\psi_4 y + x_1\lambda_4\nonumber\\
Q_3{\bf 1}&=&\psi_3 y + x_1\lambda_3\nonumber\\
Q_4{\bf 1}&=&\psi_1 y + x_1\lambda_1\nonumber\\
Q_1Q_2{\bf 1}&=&\psi_2\lambda_4-g_1 y -\psi_4\lambda_2-x_1f_2\nonumber\\
Q_1Q_3{\bf 1}&=&\psi_2\lambda_3+g_2y-\psi_3\lambda_2+x_1f_3\nonumber\\
Q_1Q_4{\bf 1}&=&\psi_2\lambda_1-{\dot x}_2y-\psi_1\lambda_2-x_1f_1\nonumber\\
Q_2Q_3{\bf 1}&=&\psi_4\lambda_3-{\dot x}_2y-\psi_3\lambda_4-x_1f_1\nonumber\\
Q_2Q_4{\bf 1}&=&\psi_4\lambda_1-g_2y-\psi_1\lambda_4-x_1f_3\nonumber\\
Q_3Q_4{\bf 1}&=&\psi_3\lambda_1-g_1y-\psi_1\lambda_3-x_1f_2\nonumber\\
Q_1Q_2Q_3{\bf 1}&=&{\dot\psi}_1y-g_1\lambda_3-g_2\lambda_4-{\psi_2}f_1-
{\dot x}_2\lambda_2-\psi_4f_3-\psi_3f_2+x_1{\dot\lambda}_1\nonumber\\
Q_1Q_2Q_4{\bf 1}&=&-{\dot\psi}_3y-g_1\lambda_1+
{\dot x}_2\lambda_4-\psi_2f_3-g_2\lambda_2+\psi_4f_1-\psi_1f_2-x_1{\dot\lambda}_3\nonumber\\
Q_1Q_3Q_4{\bf 1}&=&{\dot\psi}_4y+g_2\lambda_1+{\dot x}_2\lambda_3-\psi_2f_2-g_1\lambda_2+\psi_3f_1+\psi_1f_3+x_1{\dot\lambda}_4\nonumber\\
Q_2Q_3Q_4{\bf 1}&=&-{\dot\psi}_2y-{\dot x}_2\lambda_1+g_2\lambda_3-\psi_4f_2-g_1\lambda_4+\psi_3f_3-\psi_1f_1-x_1{\dot\lambda}_2\nonumber\\
Q_1Q_2Q_3Q_4{\bf 1}&=& -{\ddot x}_1y-x_1{\ddot y}+{\dot\psi}_1\lambda_1+{\dot\psi}_2\lambda_2+{\dot\psi}_3\lambda_3+
{\dot\psi}_4\lambda_4-\psi_1{\dot\lambda}_1-\psi_2{\dot\lambda}_2-\nonumber\\&&-\psi_3{\dot\lambda}_3-\psi_4
{\dot\lambda}_4+2{\dot x}_2f_1+2g_1f_2+2g_2f_3,
\end{eqnarray}
while in the second case the corresponding formulae are
\begin{eqnarray}
{\bf 1} &=& x_2y,\nonumber\\
Q_1{\bf 1} &=& -\psi_1y+x_2\lambda_2\nonumber\\
Q_2{\bf 1}&=&-\psi_3 y + x_2\lambda_4\nonumber\\
Q_3{\bf 1}&=&\psi_4 y + x_2\lambda_3\nonumber\\
Q_4{\bf 1}&=&\psi_2 y + x_2\lambda_1\nonumber\\
Q_1Q_2{\bf 1}&=&-\psi_1\lambda_4-g_2 y +\psi_3\lambda_2-x_2f_2\nonumber\\
Q_1Q_3{\bf 1}&=&-\psi_1\lambda_3-g_1y-\psi_4\lambda_2+x_2f_3\nonumber\\
Q_1Q_4{\bf 1}&=&-\psi_1\lambda_1+{\dot x}_1y-\psi_2\lambda_4-x_2f_1\nonumber\\
Q_2Q_3{\bf 1}&=&-\psi_3\lambda_3+{\dot x}_1y-\psi_4\lambda_4-x_2f_1\nonumber\\
Q_2Q_4{\bf 1}&=&-\psi_3\lambda_1+g_1y-\psi_2\lambda_4-x_2f_3\nonumber\\
Q_3Q_4{\bf 1}&=&\psi_4\lambda_1-g_2y-\psi_2\lambda_3-x_2f_2\nonumber\\
Q_1Q_2Q_3{\bf 1}&=&{\dot\psi}_2y-g_2\lambda_3+g_1\lambda_4+{\psi_1}f_1+
{\dot x}_1\lambda_2+\psi_3f_3-\psi_4f_2+x_2{\dot\lambda}_1\nonumber\\
Q_1Q_2Q_4{\bf 1}&=&-{\dot\psi}_4y-g_2\lambda_1-
{\dot x}_1\lambda_4+\psi_1f_3+g_1\lambda_2-\psi_3f_1-\psi_2f_2+x_2{\dot\lambda}_3\nonumber\\
Q_1Q_3Q_4{\bf 1}&=&-{\dot\psi}_3y-g_1\lambda_1-{\dot x}_1\lambda_3+\psi_1f_2-g_2\lambda_2+\psi_4f_1+\psi_2f_3+x_2{\dot\lambda}_4\nonumber\\
Q_2Q_3Q_4{\bf 1}&=&{\dot\psi}_1y+{\dot x}_1\lambda_1-g_1\lambda_3+\psi_3f_2-g_2\lambda_4+\psi_4f_3-\psi_2f_1-x_2{\dot\lambda}_2\nonumber\\
Q_1Q_2Q_3Q_4{\bf 1}&=& -{\ddot x}_2y-x_2{\ddot y}+{\dot\psi}_2\lambda_1-{\dot\psi}_1\lambda_2+{\dot\psi}_4\lambda_3-
{\dot\psi}_3\lambda_4-\psi_2{\dot\lambda}_1+\psi_1{\dot\lambda}_2-\nonumber\\&&-\psi_4{\dot\lambda}_3+\psi_3
{\dot\lambda}_4-2{\dot x}_1f_1-2g_1f_3+2g_2f_2
\end{eqnarray}
For both these enveloping representations, the reduction into its irreducible components has to be performed as in the
{\em iia}) case discussed above. 
\par{~}\par
{\em iic}) {\em The $(2,4,2)_{s=0}\times (2,4,2)_{s=0}$ case}\par
{~}\par
Let us express as
$(x_1,x_2; \psi_1,\psi_2,\psi_3,\psi_4;g_1,g_2)$ the left multiplet and the right multiplet as $(y_1,y_2; \lambda_1,\lambda_2,\lambda_3,\lambda_4;f_1,f_2)$.
At the leading (spin $s=0$) order, their tensor product produces, as {\em symmetric} (see the discussion above at the point {\em ia}) of the present appendix) bilinear multiplets, a $(2,4,2)$ irrep lead by
$x_1y_1-x_2y_2$, $x_1y_2+x_2y_1$, plus 
an enveloping $(1,4,6,4,1)$ reducible multiplet
of $N=4$ lead by $x_1y_1+x_2y_2$.
Explicitly, the bilinear multiplet $(2,4,2)\equiv ({\tilde x}_1, {\tilde x}_2;{\tilde \psi}_1,
{\tilde \psi}_2, {\tilde \psi}_3,{\tilde\psi}_4; {\tilde g}_1, {\tilde g}_2)$ is given by
\begin{eqnarray}\label{242tensor}
{\tilde x}_1&=& x_1y_1-x_2y_2\nonumber\\
{\tilde x}_2&=& x_1y_2+x_2y_1\nonumber\\
{\tilde \psi}_1&=& \psi_1y_1+x_1\lambda_1-\psi_2y_2-x_2\lambda_2\nonumber\\
{\tilde \psi}_2&=&\psi_2y_1+x_1\lambda_2+\psi_1y_2+x_2\lambda_1 \nonumber\\
{\tilde \psi}_3&=&\psi_3y_1+x_1\lambda_3-\psi_4y_2-x_2\lambda_4 \nonumber\\
{\tilde \psi}_4&=& \psi_4y_1+x_1\lambda_4+\psi_3y_2+x_2\lambda_3\nonumber\\
{\tilde g}_1&=& g_1y_1-\psi_2\lambda_4+\psi_4\lambda_2+x_1f_1-g_2y_2+\psi_1\lambda_3-\psi_3\lambda_1-x_2f_2 \nonumber\\
{\tilde g}_2&=& g_2y_1-\psi_4\lambda_1+\psi_1\lambda_4+x_1f_2+g_1y_2-\psi_3\lambda_2+\psi_2\lambda_3+x_2f_1
\end{eqnarray}
Its supersymmetric transformations are reported in Appendix {\bf A} (third case of {\em v})).
\par
The enveloping multiplet is 
\begin{eqnarray}
{\bf 1} &=&x_1y_1+x_2y_2,\nonumber\\
Q_1{\bf 1} &=& \psi_2y_1+x_1\lambda_2-\psi_1y_2-x_2\lambda_1\nonumber\\
Q_2{\bf 1}&=&\psi_4 y_1 + x_1\lambda_4-\psi_3y_2-x_2\lambda_3\nonumber\\
Q_3{\bf 1}&=&\psi_3 y_1 + x_1\lambda_3+\psi_4y_2+x_2\lambda_4\nonumber\\
Q_4{\bf 1}&=&\psi_1 y_1+ x_1\lambda_1+\psi_2y_2+x_2\lambda_2\nonumber\\
Q_1Q_2{\bf 1}&=&\psi_2\lambda_4-g_1 y_1 -\psi_4\lambda_2-x_1f_1
+\psi_1\lambda_3-g_2 y_2 -\psi_3\lambda_1-x_2f_2
\nonumber\\
Q_1Q_3{\bf 1}&=&
\psi_2\lambda_3+g_2 y_1 -\psi_3\lambda_2+x_1f_2
-\psi_1\lambda_4-g_1 y_2 +\psi_4\lambda_1-x_2f_1
\nonumber\\
Q_1Q_4{\bf 1}&=&
2\psi_2\lambda_1-2\psi_1\lambda_2 -{\dot x}_2 y_1 -x_1{\dot y}_2
+{\dot x}_1y_2+x_2{\dot y}_1
\nonumber\\
Q_2Q_3{\bf 1}&=&
2\psi_4\lambda_3-2\psi_3\lambda_4-{\dot x}_2 y_1 -x_1{\dot y}_2
+{\dot x}_1 y_2 +x_2{\dot y}_1
\nonumber\\
Q_2Q_4{\bf 1}&=&
\psi_4\lambda_1-g_2 y_1 -\psi_1\lambda_4-x_1f_2
-\psi_3\lambda_2+g_1 y_2 +\psi_2\lambda_3+x_2f_1
\nonumber\\
Q_3Q_4{\bf 1}&=&
\psi_3\lambda_1-g_1 y_1 -\psi_1\lambda_3-x_1f_1
+\psi_4\lambda_2-g_2 y_2 -\psi_2\lambda_4-x_2f_2
\nonumber\\
Q_1Q_2Q_3{\bf 1}&=&{\dot\psi}_1y_1+{\dot\psi}_2y_2+x_1{\dot\lambda}_1+x_2{\dot\lambda}_2
-\psi_1{\dot y}_1-{\psi_2}{\dot y}_2-{\dot x}_1\lambda_1-{\dot x}_2\lambda_2-\nonumber\\
&& -2g_1\lambda_3-2g_2\lambda_4-2\psi_3f_1
-2\psi_4f_2\nonumber\\
Q_1Q_2Q_4{\bf 1}&=&
-{\dot\psi}_3y_1-{\dot\psi}_4y_2-x_1{\dot\lambda}_3-x_2{\dot\lambda}_4
+\psi_3{\dot y}_1+{\psi_4}{\dot y}_2+{\dot x}_1\lambda_3+{\dot x}_2\lambda_4-\nonumber\\
&&-2g_1\lambda_1-2g_2\lambda_2-2\psi_1f_1
-2\psi_2f_2\nonumber\\
Q_1Q_3Q_4{\bf 1}&=&
-{\dot\psi}_3y_2+{\dot\psi}_4y_1-x_2{\dot\lambda}_3+x_1{\dot\lambda}_4
+\psi_3{\dot y}_2-{\psi_4}{\dot y}_1+{\dot x}_2\lambda_3-{\dot x}_1\lambda_4-\nonumber\\
&&-2g_1\lambda_2+2g_2\lambda_1-2\psi_2f_1
+2\psi_1f_2
\nonumber\\
Q_2Q_3Q_4{\bf 1}&=&
-{\dot\psi}_2y_1+{\dot\psi}_1y_2-x_1{\dot\lambda}_2+x_2{\dot\lambda}_1
-\psi_1{\dot y}_2+{\psi_2}{\dot y}_1-{\dot x}_2\lambda_1+{\dot x}_1\lambda_2-\nonumber\\
&&-2g_1\lambda_4+2g_2\lambda_3-2\psi_4f_1
+2\psi_3f_2
\nonumber\\
Q_1Q_2Q_3Q_4{\bf 1}&=& -{\ddot x}_1y_1 -{\ddot x}_2y_2
-x_1{\ddot y}_1-x_2{\ddot y}_2+2{\dot x}_1{\dot y}_1 +2 {\dot x}_2{\dot y}_2 +\nonumber\\
&&2{\dot \psi}_1\lambda_1+
2{\dot \psi}_2\lambda_2+2{\dot\psi}_3\lambda_3 +2{\dot\psi}_4\lambda_4-2\psi_1{\dot\lambda}_1-
2{\psi}_2{\dot\lambda}_2
-\nonumber\\&&-2\psi_3{\dot\lambda}_3-2{\psi}_4{\dot\lambda}_4+4g_1f_1+4g_2f_2
\end{eqnarray}
The reduction of this enveloping multiplet into its irreps constituents can be carried on as in the case
{\em iia}) examined above. 

\renewcommand{\theequation}{D.\arabic{equation}}
\setcounter{equation}{0}
\par
{~}\par
{\Large{\bf Appendix D\\{~}\\ The $N=2$ supersymmetric vacua fusion algebra}}\par
{~}\par
We present here the simplest non-trivial example of a supersymmetric vacuum fusion algebra,
giving explicit results for the $N=2$ extended supersymmetry.\par
According to the discussion of section {\bf 7}, two different $N=2$ cases can be considered. 
At first we can label the $N=2$ irreps as,
\begin{eqnarray}
\relax[1]&\equiv& (2,2)\nonumber\\
\relax [2]&\equiv& (1,2,1)
\end{eqnarray}
without distinguishing their character (bosonic or fermionic).\par
In this case it can be easily proven that the two $2\times 2$ fusion matrices $N_1$, $N_2$ are
given by
\begin{eqnarray}
N_1 &=& \left(\begin{array}{cc} $2$&$2$\\
$0$&$4$
\end{array}\right)\nonumber\\
N_2 &=& \left(\begin{array}{cc} $0$&$4$\\
$0$&$4$
\end{array}\right)
\end{eqnarray}
The two fusion matrices indeed commute, as they should do, according to the property $3$) of section ${\bf 7}$.
It is worth mentioning the usefulness of the commutativity property of the fusion algebra to explicitly
determine the fusion matrices. Already at this level in fact the fusion associated to the irrep decomposition of $[1]\times[2]$ can be written down,
without explicitly carrying out the actual computation, just by the knowledge of
the $[1]\times[1]$ and $[2]\times[2]$ fusions (which are easier to compute than the ``mixed"
$[1]\times [2]$ fusion) and of the $(2,2)\times (1,2,1) =(2,6,6,2)$ spin content 
of the reducible tensored representation.  This r.h.s. admits in principle
several decompositions into irreps. It can be easily proven, by checking the leading order terms, that
the r.h.s. is decomposed according to
$2\cdot (1,2,1)_{s=0} + (2,4,2)_{s=\frac{1}{2}}$. The $(2,4,2)_{s=\frac{1}{2}}$ representation is reducible.
It admits in principle two decompositions into irreps, either $(2,2)_{s=\frac{1}{2}}+(2,2)_{s=1}$
or $2\cdot (1,2,1)_{s=\frac{1}{2}}$. The first case, however, would produce a set of 
fusion matrices $N_1$, $N_2$ which {\em do not commute}. The second case, leading to the two commuting matrices above, is verified. \par
If we discriminate between bosonic and fermionic representations, the inequivalent $N=2$ irreps
can be labeled as follows
\begin{eqnarray}
\relax[1]&\equiv& (2,2)_{Bos}\nonumber\\
\relax [2]&\equiv& (1,2,1)_{Bos}
\nonumber\\
\relax[3]&\equiv& (2,2)_{Fer}\nonumber\\
\relax [4]&\equiv& (1,2,1)_{Fer}
\end{eqnarray}
Under this assumption we obtain an $N=2$ fusion algebra realized in terms of four $4\times 4$, mutually commuting, matrices.
They are explicitly given by
\begin{eqnarray}
N_1 &=& \left(\begin{array}{cccc} $1$&$2$&$1$&$0$\\
$0$&$2$&$0$&$2$\\
$1$&$0$&$1$&$2$\\
$0$&$2$&$0$&$2$
\end{array}\right)\nonumber\\
N_2 &=& \left(\begin{array}{cccc} $0$&$2$&$0$&$2$\\
$0$&$2$&$0$&$2$\\
$0$&$2$&$0$&$2$\\
$0$&$2$&$0$&$2$
\end{array}\right)\nonumber\\
N_3 &=& \left(\begin{array}{cccc} $1$&$0$&$1$&$2$\\
$0$&$2$&$0$&$2$\\
$1$&$2$&$1$&$0$\\
$0$&$2$&$0$&$2$
\end{array}\right)\nonumber\\
N_4 &=& N_2
\end{eqnarray}

%\end{document}
{}~
\\{}~
\par {\large{\bf Acknowledgments}}{} ~\\{}~\par
One of us (F.T.) is grateful to S. Krivonos for clarifying discussions on constrained superfields
in the extended superspace approach. M. Rojas acknowledges FAPERJ for financial support.
We are grateful to the referee for having motivated us to write the off-shell invariants of Section {\bf 7}.
%\end{document}

\end{document}